\def\ignorethis#1{{}} 

\font\ztitle=cmr10 scaled\magstep3
\font\zmsec=cmbx10 scaled 1200
\font\zssec=cmti10 scaled 1200

\def\zmsecspc{{\vskip18pt}}
\def\zssecspc{{\vskip16pt}}
\def\zmtsecspc{{\vskip14pt}}
\def\zstsecspc{\bigskip}

\baselineskip=\normalbaselineskip

\def\newpage{\vfill\eject}

\newcount\equationnumber
\newcount\sectionnumber
\sectionnumber=1
\def\zen{\the\sectionnumber.\the\equationnumber}
\def\zp{
   \global\advance\equationnumber by 1 
   (\zen)}
\def\zpa#1{
   \global\advance\equationnumber by 1 
   (\zen{{\rm #1}})}
\def\zps#1{(\zen{{\rm #1}})}
\def\advsecnum{
   \equationnumber = 0
   \global\advance\sectionnumber by 1}

\def\zseczero{}
\def\zsecone{1}
\def\zsectwo{2}
\def\zsecthree{3}
\def\zsecfour{4}
\def\zsecfive{5}

\def\a{\alpha}

\def\b{\beta}
\def\bl{\left(}
\def\br{\right)}

\def\sech{{\rm sech}}

\def\d#1{\partial_{#1}}
\def\dg{\dagger}

\def\dd{\Delta}

\def\e{\eta}
\def\ep{\epsilon}
\def\fa{{1\over\sqrt{2}}}
\def\fb{{1\over 2}}
\def\fmn{\sqrt{{M\over4\pi}}}

\def\g{\gamma}
\def\gg{\Gamma}
\def\gz{{\rm GZ}}

\def\jp{{J_\perp}}

\def\zix{{\int_{-\infty}^0 dx\,}}
\def\zit{{\int_{-\infty}^\infty dt\,}}

\def\zu{{\uparrow}}
\def\zd{{\downarrow}}

\def\l{\lambda}
\def\ll{\Lambda}
\def\lt{{\tilde\lambda}}
\def\lc{\Lambda}

\def\pt{{\tilde \phi}}
\def\ph{{\hat \phi}}
\def\ppt{{\tilde \psi}}
\def\p#1{\psi_{#1}}
\def\pb#1{{\overline{\psi}}_{#1}}
\def\pp{\psi_+}
\def\ppm{\psi_-}
\def\pbp{{\overline\psi}_+}
\def\pbm{{\overline\psi}_-}

\def\r{\rho}
\def\rt{{\tilde r}}

\def\rb#1{{ |#1\rangle}}

\def\t{\theta}
\def\s{\sigma}
\def\sp{{d^\dagger}}
\def\sm{d}
\def\sgc{G}
\def\w{\omega}
\def\wb{{\overline\omega}}

\def\vp{\varphi}
\def\vpb{{\overline\varphi}}
\def\vt{\vartheta}
\def\vk{{\vec k}}
\def\vx{{\vec x}}

\def\ukondo{13}
\def\Saleur{14}
\def\Bassi{15}
\def\Warner{16}
\def\Skorik{17}
\def\Ghoshal{18}
\def\Corrigan{19}
\def\Leclair{20}
\def\Ameduri{21}
\def\Ginsparg{22}
\def\Coleman{23}
\def\Mandel{24}
\def\Bernard{25}
\def\Wieg{26}   
\def\Fendley{27}
\def\Fring{28}
\def\Grad{29}
\def\Zam{30}
\def\Aludwig{31}
\def\Bowcock{32}
\def\Legett{33}
\def\Guinea{34}
\def\Schotte{35}
\def\Delft{36}

\def\Emery{39}

\baselineskip=1.5\normalbaselineskip

\ignorethis{\magnification 1200}
\headline={\hfil CLNS 98/1592; hep-th/9811138}
\footline={7/99\hfil}

\topskip=0.7in
\centerline{{\ztitle The Kondo Model with a Bulk Mass Term}}
\vskip0.8in
\centerline{Zorawar S. Bassi$^\dg$}
\centerline{Andr\'e LeClair}
\bigskip
\centerline{Newman Laboratory}
\centerline{Cornell University}
\centerline{Ithaca, NY 14853, USA}
\vskip0.5in

\noindent
We introduce two massive versions of the anisotropic spin 1/2 Kondo model
and discuss their integrability.  The two models have the
same bulk sine-Gordon interactions, but  differ in their boundary
interactions. 
At the  Toulouse free fermion point each of the models 
can be understood  as  two decoupled Ising models in boundary magnetic
fields.
Reflection S-matrices away from the free fermion point are conjectured.

\vskip3.85in
\noindent
$^\dg$ zsb1@cornell.edu

\newpage


\advance\pageno by -1
\headline={\hfil}
\footline={\hfil\folio\hfil}

\topskip=10pt

\noindent
{\zmsec \zseczero Introduction}\par
\nobreak
\zmtsecspc
The s-d exchange model, often referred to as the Kondo model, has been
extensively discussed since it was first used by Kondo [1] to calculate 
the effects of impurities on the resistivity of a metal.  Various approaches
have been applied to study this model, including renormalization group
methods [2], the Bethe ansatz [3]-[9] and conformal field theory [10]-[12].  
Traditionally
the Kondo model contains a single coupling $J$.  Allowing the exchange 
couplings to differ gives the anisotropic Kondo model.  Regardless of whether
there is an anisotropy or not, the Kondo model is a massless integrable
system for arbitrary spin [\ukondo],[\Saleur].

In this paper we present two massive generalizations of the anisotropic
spin 1/2 Kondo model and discuss their integrability.  These models are
bosonic boundary field theories related to the original fermionic system 
via bosonization.  Such massive theories have possible applications to
1D impurity problems with an excitation gap.  An example is 
a magnetic  impurity in a 
superconductor, where the mass  represents the BCS energy gap [\Bassi].

The motivation for considering our particular models came as follows. 
The massive boundary sine-Gordon model (MBSG) is a boundary
field theory of a single boson $\phi$ with a  bulk interaction  in the
action 
$G \int dx dt \cos( \beta \phi)$ and a boundary interaction 
$\lambda \int dt \cos (\beta (\phi- \phi_0 ) /2 )$.   
The model is known to be integrable
for arbitrary $G, \lambda$ and $\phi_0$.  The massless limit $G\to 0$ is
well defined, which defines the massless boundary sine-Gordon model [\Warner].
On the other hand, the usual anisotropic Kondo model after bosonization
has a boundary sine-Gordon like interaction but with the inclusion 
of spin operators $S_\pm$ at the boundary:  
$ \lambda\int dt  ~ S_+ e^{i\beta \phi/2 } + S_- e^{-i\beta \phi/2} $, and
the bulk is just a free massless scalar field. If one compares the  massless
boundary sine-Gordon  with the Kondo model, one finds that the
same sine-Gordon spectrum of particles diagonalizes the boundary interaction;
the difference is in the detailed form of the scattering matrices for
reflection off the boundary [\Skorik]. 
  So the question naturally arises whether
it is possible to add a bulk sine-Gordon term to the Kondo theory
and preserve the integrability, in analogy with the massive boundary
sine-Gordon theory.  

If the theories we define in this paper are indeed integrable this
implies that the usual bulk sine-Gordon integrals of motion are
not spoiled by the boundary interactions.  In this work we do not
establish the integrability of our models by studying these integrals
of motion. 
Rather we take the following approach to studying the problem.  In 
[\Ghoshal] the most general solution to the boundary
Yang-Baxter equation corresponding to a massive sine-Gordon spectrum
and no boundary degrees of freedom was described and related to the
massive boundary sine-Gordon model.  The known scattering
description of the Kondo 
model also does not require boundary degrees of freedom for spin $1/2$
due to screening.  In the massive theory one might expect a competition
between the gap effects and the screening, since the screened state
is a property of a new infra-red fixed point, but the existence
of the gap introduces an infra-red cutoff, and this
 may imply that this infra-red fixed point is never
reached.  However non-trivial renormalization group beta
functions only occur at the isotropic point.  Our analysis of
the free fermion point indeed indicates that a variant of  screening does
occur in the massive case.      We  thus 
assume this continues to hold for the massive
model, namely that the scattering description does not require any
boundary degrees of freedom, at least in some sector. 
Then if the massive version of the Kondo model is integrable,
since it involves a bulk sine-Gordon theory, its reflection S-matrix
off the boundary must be contained in the general solution found by
Ghoshal and Zamolodchikov (GZ) [\Ghoshal] up to overall scalar (CDD)
 factors.   If one simply
adds a bulk sine-Gordon term to the usual Kondo model, and specializes
to the free fermion point, then the reflection amplitudes can
be computed explicitly and it is found that they do not
coincide with those of GZ, even up to CDD factors.  
However we found that a slight 
modification
of the boundary interaction does match onto the GZ solution    
and it is this model for which we conjecture an extension away from
the free fermion point. This latter model requires that
the boundary and bulk couplings not be independent.
The first model on the other hand does
not have the constraints of the boundary Yang-Baxter equation at
the so-called reflectionless points, so in this situation we  also
conjecture an exact reflection S-matrix.   

In the massless cases the main distinction between the Kondo and
boundary sine-Gordon theories is that sine-Gordon has a flow between
free boundary conditions in the ultra-violet  and fixed boundary conditions
in the infra-red, whereas the Kondo model maintains a free boundary 
condition throughout. 
Our conjectured reflection S-matrix for the modified
massive Kondo model is in accordance with this, in that it corresponds to
the reflection S-matrix for the boundary sine-Gordon theory at
the free boundary condition times a CDD factor.

We present our results as follows. 
In section \zsecone\ we define  the two  massive 
versions of the Kondo model that we consider.  The first,
simply referred to as massive Kondo (MK), is just a sine-Gordon model
with the usual Kondo interaction at the boundary.  The second,
the modified massive Kondo model (mMK), has a slightly different
boundary interaction.  More importantly, in the  mMK model 
the boundary coupling $\lambda$ is not independent of the bulk
coupling $G$, but satisfies $G \propto \lambda^2$. 
This is reminiscent of what happens in the 
boundary Toda theories [\Corrigan].
In section \zsectwo\ we study both models
at the free fermion point, and relate them both to two decoupled
Ising models in appropriate boundary magnetic fields.  In section \zsecthree\
we compare with the GZ solution and propose reflection S-matrices 
away from the free fermion point.  Section 4 contains our results
on boundary bound states for the massive models.    
An appendix reviews the mapping of the Kondo model to a bosonic boundary
field theory. 



\zmsecspc
\noindent
{\zmsec \zsecone. The Field Theory Models}\par
\nobreak
\zmtsecspc
The anisotropic Kondo Hamiltonian is
$$ H^K = \sum_{\vk\s} \ep(\vk) c_{\vk\s}^\dg c_{\vk\s} +
{J_z\over 2} s_z \sum_{\vk\vk^\prime\s{\s^\prime}} c_{\vk\s}^\dg
{(\s^z)}_{\s{\s^\prime}}  c_{{\vk^\prime}{\s^\prime}} +
{\jp\over 2} \sum_{\vk\vk^\prime\s{\s^\prime}}\left(  s_x c_{\vk\s}^\dg
{(\s^x)}_{\s{\s^\prime}}  c_{{\vk^\prime}{\s^\prime}} + s_y c_{\vk\s}^\dg
{(\s^y)}_{\s{\s^\prime}}  c_{{\vk^\prime}{\s^\prime}} \right),
\eqno\zp $$
\edef\KorigHZ{\zen}%
where $c_{\vk\s}^\dg$ are conduction electron creation operators with wave
vector $\vk$ and spin $\s$ (up $\zu$ or down $\zd$), $\s^{x,y,z}$ are the 
Pauli matrices, $\vec s$ is the impurity spin operator (spin 1/2)
at $\vec r=0$ and $J_z$ and $\jp$ are the exchange couplings.  In the 
isotropic Kondo model $J_z = \jp$.  For s-wave scattering
near the fermi surface, the spin sector of (\KorigHZ) can be mapped onto
the following bosonic boundary field theory 
$$ H^K = \fb\zix \bl (\d t \phi)^2 + (\d x \phi)^2\br + \l \bl
S_+ e^{i\b\phi(0)/2} + S_- e^{-i\b\phi(0)/2} \br,
\eqno\zp$$
\edef\KhamZ{\zen}%
where $\l$ is the coupling and $\b$ determines the anisotropy.  
The isotropic point corresponds to $\b=\sqrt{8\pi}$.
The series of steps leading to (\KhamZ) are outlined in the appendix.

The boundary field theory
 (\KhamZ) is our starting point, which  we will simply 
refer to as the Kondo model.  The parameters $\l$ and $\b$ are 
considered to be arbitrary parameters, independent of their relation 
to the couplings $J_z$ and $\jp$ (see appendix).  
This model is known to be integrable for 
any spin.  However, integrability requires that the matrices $S_i$ form a
spin $j$ representation of the $q$-deformed quantum algebra
$su(2)_q$ [\ukondo], where $q = \exp (i\beta^2/8)$.  
The $su(2)_q$ relations are
$$ [S_z,S_\pm] = \pm 2 S_\pm,\quad [S_+,S_-] =
{q^{S_z} - q^{-S_z}\over q - q^{-1}}.\eqno\zp $$
\edef\KsuqZ{\zen}%
For the isotropic case, $q=-1$ and (\KsuqZ) reduces to the usual $su(2)$
algebra.  For spin 1/2, the difference between $su(2)_q$ and $su(2)$ is
not important since the Pauli matrices also form a representation of 
$su(2)_q$ for arbitrary $q$.

\zssecspc
\noindent
{\zssec \zsecone.1. The Massive Kondo Models}\par
\nobreak
\zstsecspc
We are interested in a massive generalization of (\KhamZ), which 
in the massless limit reproduces the Kondo behaviour.  In particular, the
massless limit should give the spin 1/2 Kondo scattering matrix.  This has
led us to consider two models.

The first model, the MK model, has the Hamiltonian
$$ H^{MK} = \fb\zix \bl (\d t \phi)^2 + (\d x \phi)^2 - 
{\sgc} \cos(\b\phi) \br + \l \bl
S_+ e^{i\b\phi(0)/2} + S_- e^{-i\b\phi(0)/2} \br. \eqno\zp$$
\edef\MKhamZ{\zen}%
The bulk part is simply the sine-Gordon model.  This
model was studied in [\Leclair] at the free fermion 
point $(\b=\sqrt{4\pi})$ and 
the reflection S-matrix was calculated.  We will re-derive the free
fermion reflection amplitudes and discuss their extension to arbitrary $\b$.
It is known that the bulk sine-Gordon spectrum also diagonalizes the
boundary interaction [\Ghoshal], thus (\MKhamZ) is perhaps 
the most obvious massive
generalization.  The bulk spectrum consists of solitons for 
$\sqrt{4\pi}\leq \b<\sqrt{8 \pi}$, and for $\b< \sqrt{4\pi}$ there are also
breathers.  (For $\b>\sqrt{8\pi}$ the bulk energy is unbounded.)  The
soliton and breather masses are functions of $\sgc$ and $\b$. 
We will argue that unlike the Kondo model, (\MKhamZ) is not integrable for 
arbitrary $\b$, but only at certain so-called reflectionless points.

This brings us to the second model, the mMK model, with the Hamiltonian
$$\eqalignno{ H^{mMK} = \fb\zix \Bigl( (\d t \phi)^2 + (\d x \phi)^2 -
& \,{\sgc} \cos(\b\phi) \Bigr) &\cr 
- i \l & \bl
\cos{\b\over 2} (\phi(0)-\ph) S_+ - S_- \cos{\b\over 2}(\phi(0)-\ph^*)\br,
& \zpa{a} \cr }$$
where
$$ \ph = - {2\over \b} \bl {\pi\over 2} - i\ph_0\br, \eqno\zps{b} $$
and the coupling $\l$ is related to the bulk coupling $\sgc$
$$ \l^2 \propto \sgc,\quad \l = \sqrt{2 M}\ {\rm at\ } \b=\sqrt{4\pi}.
\eqno\zps{c} $$
\edef\mMKhamZ{\zen}%
Here $\ph_0$ is an arbitrary real parameter and $M$ is the fermion
(soliton)  mass.
\ignorethis{The cutoff $a^{-1/2}$ is included to give the correct dimensions
and for regularizing the re-fermionization of the interaction (see below).}
Let us compare this with the MBSG model 
$$ H^{MBSG} = \fb\zix \bl (\d t \phi)^2 + (\d x \phi)^2 - 
{\sgc} \cos(\b\phi) \br + \lambda 
\cos{\b\over 2} (\phi(0)-\phi_0). \eqno\zp$$
\edef\MBSGhamZ{\zen}%
The MBSG model is integrable for all values of $\lambda$,  $\phi_0$
and $G$, and all can be considered as independent [\Ghoshal].
In the mMK
model there is only one free coupling $\ph_0$, with $\l$ 
being a function of the bulk coupling $\sgc$, i.e. not a free parameter.  
We believe that
the mMK model is integrable for all values of $\b$ and conjecture  
a soliton reflection S-matrix in section 3, which in   
the massless limit  recovers the
Kondo model.  Note that for both massive models, the scaling dimension of
the bulk coupling $\sgc$ is twice that of the boundary coupling $\l$.  
Because of the cosines and exponentials, these couplings 
have anomalous dimensions, resulting in a scaling 
dimension of $2- \b^2/4\pi$ for $\sgc$ and $1 - \b^2/8\pi$ for $\l$.

\advsecnum

\zmsecspc
\noindent
{\zmsec \zsectwo. Boundary Reflection S-Matrices at the Free Fermion Point}\par
\nobreak
\zmtsecspc
In this section  we derive the reflection S-matrices at the free fermion 
point starting from the action.
A similar procedure was used in [\Ameduri] for the MBSG model.  
For the MK model
this was first done in [\Leclair].  Here we repeat that calculation in
a more general context, and by making a change of basis give a clearer
understanding into the structure of the reflection S-matrix.

The bulk action for both models is
$$ S_{\rm bulk} = \fb \zit\zix \bl (\d t\phi)^2 -(\d x \phi)^2 +
{\sgc} \cos(\b\phi)\br. \eqno\zp $$
\edef\BblkactZ{\zen}%
In the massless limit, the scalar field can be separated into its
right and left moving components
$$ \phi(z^+,z^-) = \vp(z^+) + \vpb(z^-), \quad z^\pm = t\pm x, \eqno\zp $$
in terms of which the Neumann boundary condition $\d x\phi=0$ at $x=0$ 
becomes
$$ \d {z^+} \vp(z^+) = \d {z^-} \vpb(z^-),\quad (x=0).\eqno\zp$$
This implies
$$ \vp(z^+) = \vpb(z^-) - {\s\over \sqrt{4\pi}},\quad (x=0),
\eqno\zp$$
\edef\BbcZ{\zen}%
where $\s$ is a constant related to the zero modes of $\phi$ [\Ameduri],
[\Ginsparg].  The massless boson bulk theory is equivalent to a
massless Dirac theory through bosonization/fermionization.  Let 
$\p {\pm}(z^+)$ and $\pb {\pm}(z^-)$ respectively be the left and right 
chiral components of the Dirac fermion with $U(1)$ charge 
$\pm 1$. Then the bosonization relations read
$$ \p {\pm}(z^+) = e^{\pm i\sqrt{4\pi}\vp(z^+)} , \quad
\pb {\pm}(z^-) = e^{\mp i\sqrt{4\pi} \vpb(z^-)},\quad
\p - = (\p +)^\dg,\ \pb - = (\pb +)^\dg. \eqno\zp$$
\edef\bfrcZ{\zen}%
In terms of the fermions the (free) boundary  condition becomes 
$$ \p \pm = e^{\mp i\s} \pb \mp,\quad (x=0) \eqno\zp $$
\edef\fbcZ{\zen}%
which breaks the charge symmetry.

For the massive case, the bulk theory (\BblkactZ) is 
equivalent to a free massive
Dirac theory at $\b =\sqrt{4\pi}$ [\Coleman],[\Mandel].  
The fields $(\p \pm,\pb \pm )$ are now no longer chiral.  These can still
be related to $\phi$ as in (\bfrcZ), but now $\vp$ and $\vpb$, 
with $\phi=\vp+\vpb$, are not simply the left and 
right moving components [\Bernard].
The fermions $\p \pm$ and $\pb \pm$ correspond to the solitons in the
sine-Gordon model.  
Considering the massive system as a perturbation of the
massless one, we use the same boundary conditions (\BbcZ) and (\fbcZ).  The
massive fermionic action which enforces (\fbcZ) takes the form
$$ S = S_{\rm bulk} + S_{\rm bc}  \eqno\zp $$
\edef\asactniZ{\zen}%
$$ S_{\rm bulk} = S_{\rm kin} + S_{\rm mass} \eqno\zpa{a} $$
$$ S_{\rm kin} = \int_{-\infty}^\infty\,dt \int_{-\infty}^0\,dx\,
\bigl[ i\p - (\d t - \d x)\p + + i\p +(\d t - \d x)\p - +
i\pb - (\d t + \d x)\pb + + i\pb + (\d t + \d x)\pb - \bigr]\eqno\zps{b} $$
$$ S_{\rm mass} = 2i \int_{-\infty}^\infty\,dt \int_{-\infty}^0\,dx\,
 M\bigl(\p - \pb + - \pb - \p +  \bigr) \eqno\zps{c}$$
\edef\gmtZ{\zen}%
$$ S_{\rm bc} = -i \int_{-\infty}^\infty\,dt\,\bigl( e^{i\s} \p + \pb + +
 e^{-i\s} \p - \pb -\bigr), \eqno\zp $$
where $M$ is the soliton mass. 
The action $S_{\rm bulk}$ is simply a  Dirac action.   
$S_{\rm bc}$ is required to produce the boundary conditions.  
In varying $S$ with respect to all the fields, the 
boundary terms give (\fbcZ).
Note that the $\psi$'s are different from the 
original fermions occurring in the Kondo model.  
We went from (\KorigHZ) to the bosonic boundary field theory for  
$\phi$ (\KhamZ) by bosonizing (see appendix), 
and now we have (i) added a mass term and
(ii) re-fermionized, going from $\phi$ to a new set of Dirac fermions.  
The mass term (\gmtZ c) is not a mass term for the original Kondo fermions.
(In the Kondo model a similar transformation 
gives the resonant model [\Wieg],[\Legett].)
The action (\asactniZ) is common to both the MK and mMK model.

We now specify the interaction with the spin operators $S_\pm$. 
It will be convenient to consider a more
general boundary interaction, with the MK and mMK interactions being special
cases.  Consider the following action 
$$ S_{\rm int} = - {\l_1}\int_{-\infty}^\infty\,dt\ \left(
e^{i\b\phi(0)/2} S_+ + S_- e^{-i\b\phi(0)/2}\right)
+ {\l_2} \int_{-\infty}^\infty\,dt\ \left(
e^{-i\b\phi(0)/2} S_+ + S_- e^{i\b\phi(0)/2}\right),\eqno\zp $$
\edef\genmkZ{\zen}%
where $\l_1$ and $\l_2$ are couplings which can be different.  The MK 
model is obtained by setting $\l_2 = 0$ and   
$\l_1 = \l$.   For the mMK model we need to take 
$\l_1 = \l e^{\ph_0}$ and $\l_2 = \l e^{-\ph_0}$.
Using (\BbcZ) and (\bfrcZ) the interaction term (\genmkZ) can be written as
$$\eqalignno{ S_{\rm int} = - {\l_1 \over 2}\int_{-\infty}^\infty\,dt\bigl[ 
(\pp a_-  + & \pbm a_+) S_+ + S_-(a_+\ppm + a_-\pbp)\bigr] &\cr
& + {\l_2 \over 2}\int_{-\infty}^\infty\,dt\bigl[ 
(a_+\ppm + a_-\pbp) S_+ + S_- (\pp a_-  + \pbm a_+)\bigr],
& \zp \cr} $$
\edef\genmkffZ{\zen}%
where $a_\pm$ are the zero mode contributions 
$$ a_+ = e^{-i\s/2},\quad a_- = (a_+)^\dg = e^{i\s/2}, \eqno\zp $$
and we have taken the average in fermionizing
$$  e^{+i\sqrt{\pi}\phi(0)} = \fb(\p + a_- + \pb - a_+),
\quad e^{-i\sqrt{\pi}\phi(0)} = \fb(a_+ \p - + a_- \pb +).
\eqno\zp$$
In order to ensure that the interaction is a bosonic scalar, the $a_\pm$
need to be treated as anticommuting with the $\psi$'s
$$ a_\pm \psi = - \psi a_\pm. \eqno\zp $$
\edef\apropZ{\zen}%
Aside from (\apropZ), they should be regarded 
as ordinary complex numbers with
$a_\pm^2 = e^{\mp i\s},\ a_+a_- = a_-a_+ = 1.$  We can rewrite (\genmkffZ) 
in a slightly different form.  Rearranging we have
$$\eqalignno{ S_{\rm int} = - {\l_1 \over 2}\int_{-\infty}^\infty\,dt\bigl[ 
(\pp e^{i\s}  + & \pbm ) a_+ S_+ + S_- a_-(\ppm e^{-i\s} + \pbp)\bigr] &\cr
&    - {\l_2 \over 2}\int_{-\infty}^\infty\,dt\bigl[ 
(\ppm + \pbp e^{i\s}) a_+ S_+ + S_- a_-(\pp   + \pbm e^{-i\s})\bigr].
& \zp\cr} $$ 
Define the operators $\sm$ and $\sp$
$$ \sm = S_- a_-, \quad \sp = a_+ S_+,\eqno\zp$$
\edef\SrdZ{\zen}%
which are to be treated as anticommuting with 
the $\p{}$'s.  The fermionized version of (\genmkZ) becomes
$$\eqalignno{ S_{\rm int} = - {\l_1 \over 2}\int_{-\infty}^\infty\,dt\bigl[ 
(\pp e^{i\s}  +& \pbm ) \sp + \sm (\ppm e^{-i\s} + \pbp)\bigr] &\cr
& - {\l_2 \over 2}\int_{-\infty}^\infty\,dt\bigl[ 
(\ppm + \pbp e^{i\s}) \sp + \sm (\pp   + \pbm e^{-i\s})\bigr].
& \zp \cr}  $$
This is the form of the interaction we will use to calculate the reflection
S-matrices.  We want to calculate the reflection S-matrices 
directly from the boundary equations of motion (see below).  The boundary
equations of motion can be obtained by varying the action with
respect to all the fields, including $\sm$ and $\sp$, which are dynamic
variables, $\sm=\sm(t),\ \sp=\sp(t)$.  In order to obtain the correct 
boundary equations for these operators it is necessary to introduce the
following kinetic term  
$$ S_{\rm d-free} = i \int_{-\infty}^{\infty}\, dt\, \sp(t)\d t \sm(t).
\eqno\zp $$
\edef\SactZ{\zen}%
The action which we vary is thus
$$ S = S_{\rm bulk} + S_{{\rm bc}} +  S_{\rm d-free} + S_{\rm int}.\eqno\zp $$
\edef\actionZ{\zen}%

The term (\SactZ) needs some explanation.
In writing (\SactZ) we have considered $\sm(t)$ and $\sp(t)$ to be one
dimensional ``fermionic" fields.
On the other hand,  (\SrdZ) would imply that $\sm$ and $\sp$ are
$su(2)$ spin operators.
In the isotropic Kondo model it is well known that there
is a screening effect [\Fendley], whereby the spin $j$ operators $S_i$ 
act in the spin $j - 1/2$ representation.  We claim that a similar
``screening" effect also occurs in the anisotropic massive models.
This implies that for spin 1/2 the $su(2)$ structure
is lost.  The operators $\sm$ and $\sp$ are thus screened versions 
of the spin operators $S_\pm$, 
which for spin 1/2 can be effectively treated as one dimensional 
fermions satisfying
$$ {\sm}^2 = {\sp}^2 = 0,\quad \{\sm,\sp \} = 1, \eqno\zp $$
\edef\asdalgZ{\zen}%
and due to (\SrdZ) anticommuting with the $\psi$'s
$$ \sm \psi = - \psi \sm,\quad \sp \psi = - \psi \sp . \eqno\zp $$
\edef\asdapsiZ{\zen}%
Equations (\SrdZ) and (\SactZ) are statements of this screening.  
Further justification for (\SactZ) comes from the equations of motion for
for $S_- a_-$ and $a_+ S_+$.  The equations of motion can be calculated in
the Heisenberg representation, namely $\partial_t (S_- a_-) = i[H, S_- a_-]$
and $\partial_t(a_+ S_+) = i[H, a_+ S_+]$, by making use of (\asdalgZ) and
(\asdapsiZ).  This leads to the same equations of motion as from the above
Lagrangian formulation in terms of $d$ and $d^\dagger$, hence giving the
same scattering matrix.  
Thus instead of 
introducing (\SrdZ) and (\SactZ), we could have used the Hamiltonian to obtain 
$\partial_t(S_- a_-)$ and $\partial_t(a_+ S_+)$, as was done in [\Leclair].
In short, (\SactZ) is consistent with a screening effect and leads to the
correct equations of motion.  A term similar to (\SactZ) is 
also introduced in the
MBSG model at the free fermion point [\Ameduri] and the boundary Ising model
[\Ghoshal] to describe the dynamics of the boundary operators.


Varying (\actionZ) with respect to 
all the fields, $( \p \pm,\pb \pm,\sm,\sp )$,
and requiring the boundary terms to vanish, we obtain six equations 
(at $x=0$)
$$\pp - \pbm e^{-i\s} = {i\over 2}(\l_1 e^{-i\s} \sm - \l_2 \sp)\eqno\zpa{a}$$
$$\ppm - \pbp e^{i\s} = -{i\over 2}(\l_1 e^{i\s} \sp - \l_2 \sm)\eqno\zps{b}$$
\edef\voneZ{\zen}%
$$\pp e^{i\s} - \pbm = {i\over 2}(\l_1 \sm - \l_2 e^{i\s} \sp)\eqno\zpa{a}$$
$$\ppm e^{-i\s}- \pbp = -{i\over 2}(\l_1 \sp - \l_2 e^{i\s} \sm)\eqno\zps{b}$$
\edef\vtwoZ{\zen}%
$$ i\d t \sm = -{\l_1\over 2}(\pp e^{i\s} + \pbm) -
{\l_2 \over 2} (\ppm + \pbp e^{i\s} ) \eqno\zpa{a} $$
$$ i\d t \sp = {\l_1\over 2}(\ppm e^{-i\s} + \pbp) +
{\l_2 \over 2} (\pp + \pbm e^{-i\s} ). \eqno\zps{b} $$
\edef\vthreeZ{\zen}%
We see that (\voneZ) and (\vtwoZ) are identical, and (\voneZ a) and (\voneZ b) 
are adjoints, as are (\vthreeZ a) and (\vthreeZ b).  
Taking time derivatives of (\voneZ) and using (\vthreeZ),
we get the boundary equations of motion
$$ \d t\left( \pp - e^{-i\s} \pbm\right) =
- {1\over 4} (\l_1^2 + \l_2^2) \left(\pp + e^{-i\s}\pbm\right)
- {1\over 2} \l_1 \l_2 \left(\ppm e^{-i\s} + \pbp\right)\eqno\zpa{a} $$
$$ \d t\left( \ppm - e^{i\s} \pbp\right) =
- {1\over 4} (\l_1^2 + \l_2^2) \left(\ppm + e^{i\s}\pbp\right)
- {1\over 2} \l_1 \l_2 \left(\pp e^{i\s} + \pbm\right). \eqno\zps{b} $$
\edef\FbdyeomZ{\zen}%
The bulk variation of $S_{\rm bulk}$ gives the bulk equations of motion
$$ (\d t - \d x)\p \pm + M \pb \pm = 0, \quad
(\d t + \d x)\pb \pm - M \p \pm = 0. \eqno\zp $$
\edef\FblkeomZ{\zen}%
Mode expansions satisfying (\FblkeomZ) take the form
$$ \pp  = \fmn\int_{-\infty}^\infty d\t\, 
e^{-\t/2}\bigl( \wb A_-(\t) e^{-i\vk\cdot \vx}
+\w A_+^\dg(\t) e^{i\vk\cdot \vx} \bigr) \eqno\zpa{a}$$
$$ \pbp = \fmn\int_{-\infty}^\infty d\t\, 
e^{\t/2}\bigl( \w A_-(\t) e^{-i\vk\cdot \vx}
+\wb A_+^\dg(\t) e^{i\vk\cdot \vx} \bigr) \eqno\zps{b}$$
$$ \ppm = (\pp)^\dg,\quad \pbm = (\pbp)^\dg, \eqno\zps{c}$$
\edef\FmeZ{\zen}%
where $\w = \exp(i\pi/4)$, $\wb = \exp(-i\pi/4)$ and 
$\vk\cdot \vx = M t\cosh\t - M x\sinh\t$.  The variable $\t$ is the 
rapidity and $A_\pm^\dg (A_\pm)$ are fermion creation (annihilation)
operators satisfying
$$ \{ A_\pm(\t), A_\pm^\dg (\t^\prime)\} = \delta(\t-\t^\prime).\eqno\zp$$
\edef\AcomZ{\zen}%

The boundary scattering of the particles at $x=0$ can be formulated by
introducing a formal boundary operator $B$, and interpreting the boundary
equations (\FbdyeomZ) to hold when acting on $B$ [\Ghoshal].  
This will give us a set of
equations, one equation for each type of particle $A_a^\dg(\t)$, 
of the form (sum over $b$)
$$ A_a^\dg(\t)B = R_a^b(\t) A_b^\dg(-\t)B, \eqno\zp $$
\edef\RSmeqZ{\zen}%
where $R(\t)$ is called the reflection S-matrix, with $R_a^b(\t)$ 
being the amplitude for
an incoming particle $a$ to be reflected as an outgoing particle $b$.
If the boundary has additional structure allowing it to exist in 
several stable degenerate states, then (\RSmeqZ) needs to be generalized to
$$ A_a^\dg(\t)B_\alpha = R_{a\alpha}^{b\b}(\t) A_b^\dg(-\t)B_\b, 
\eqno\zp $$ 
\edef\RSmbbeqZ{\zen}%
where $\{B_\alpha\}$ are the boundary states.  The reflection S-matrix
element $R_{a\alpha}^{b\b}(\t)$ gives the amplitude for the process 
$(a,\alpha) \rightarrow (b,\b)$.  Based on the ``screening'' discussed
above, we 
will assume that (\RSmeqZ) is sufficient to describe the boundary scattering 
involving a spin 1/2 impurity.  It is known that this is sufficient
in the massless Kondo case.   However this does not entirely exclude
the possible existence of boundary structure $B_\alpha$, as it may
turn out that certain identifications of the matrix 
elements $R_{a\alpha}^{b\beta}$ reduce the problem to (\RSmeqZ).     
That is to say, the amplitudes
$R_{a}^{b}$ can describe the scattering process even if there is
some (degenerate) boundary structure.  Various 
symmetries of the system can effectively
reduce the number of independent amplitudes $R_{a\a}^{b\b}$ to a smaller set
satisfying the same scattering constraints as $R_a^b$.   
Such a situation will
occur if the soliton transition $a\rightarrow b$ restricts the boundary
transition $\a\rightarrow\b$ to be of a specific type, with all amplitudes 
having the same soliton transition being equal.  In this case the indices
$(\a,\b)$ become redundant since the amplitudes 
for $(a,\a)\rightarrow (b,\b)$ are uniquely classified by the soliton
quantum numbers $(a,b)$, i.e. $R_{a \a}^{b \b}$ reduces to $R_a^b$, with
any boundary structure being implicitly incorporated in $R_a^b$.
Poles of $R_a^b(\t)$ may then be indicative of this ``hidden'' boundary
structure.  As a specific example, suppose the boundary is described by
two degenerate states $B_\pm$ with charge $(+,-)$ as for the solitons.  
Furthermore, let the symmetries (or other constraints) of the system 
be such that (i) every
soliton charge (non-)conserving transition is accompanied by
a boundary charge (non-)conserving transition and (ii) charge conjugation
symmetry holds separately in the 
soliton and boundary sectors.  Then $R_{a\a}^{b\b}$
reduces to only two independent amplitudes of the form $R_{a \a}^{a\a}$ and
$R_{a\a}^{{\overline a}{\overline \a}}$, where the bar denotes the charge
conjugated state.  The scattering process is thus effectively described by 
$R_a^a \equiv R_{a \a}^{a\a} $ and 
$R_a^{\overline a} \equiv R_{a\a}^{{\overline a}{\overline \a}}$. 
(A similar reduction occurs in the bulk S-matrix at the reflectionless
points.)
There is also the  possibility of excited ``boundary bound states," 
arising from
poles in the reflection 
amplitudes $R_a^b(\t)$ (see section 4).  
These issues will be considered in the sequel.  

For the solitons (\RSmeqZ) reads
$$  A_\pm^\dg(\t) B = R_\pm^\pm(\t) A_\pm^\dg(\t) B + 
R_\pm^\mp(\t)A_\mp^\dg(\t) B. \eqno\zp $$
\edef\RSmseqZ{\zen}%
In the case of higher spin, $j \geq 1$, it is likely that
one needs to consider (\RSmbbeqZ), with the $\a$ index 
taking into account any
additional boundary structure due to the $S_\pm$ operators.


We now  calculate the reflection amplitudes.  
Substituting the mode expansions (\FmeZ) into
(\FbdyeomZ) gives
$$\eqalignno{
\w e^{-\t/2} \biggl( i M\cosh\t + {1\over 4} (\l_1^2 + \l_2^2) -
{i\over 2} \l_1 \l_2 e^\t \biggr) A_+^\dg(\t)& \cr 
 - \wb e^{-i\s} e^{\t/2} 
\biggl( i M\cosh\t & - {1\over 4} (\l_1^2 + \l_2^2) -
{i\over 2} \l_1 \l_2 e^{-\t} \biggr) A_-^\dg(\t) \cr
 = - \w e^{\t/2} \biggl(i M\cosh\t + {1\over 4} (\l_1^2 + \l_2^2) -
{i\over 2} \l_1 \l_2 & e^{-\t} \biggr) A_+^\dg(-\t) \cr
+ \wb e^{-i\s} e^{-\t/2} 
\biggl( i M &\cosh\t  - {1\over 4}(\l_1^2 + \l_2^2) - 
{i\over 2} \l_1\l_2 e^\t \biggr) A_-^\dg(-\t) & \zpa{a}\cr}$$
$$\eqalignno{
\w e^{-\t/2} \biggl( i M\cosh\t + {1\over 4} (\l_1^2 + \l_2^2) -
{i\over 2} \l_1 \l_2 e^\t \biggr) A_-^\dg(\t)& \cr 
 - \wb e^{i\s} e^{\t/2} 
\biggl( i M\cosh\t & - {1\over 4} (\l_1^2 + \l_2^2) -
{i\over 2} \l_1 \l_2 e^{-\t} \biggr) A_+^\dg(\t) \cr
 = - \w e^{\t/2} \biggl(i M\cosh\t + {1\over 4} (\l_1^2 + \l_2^2) -
{i\over 2} \l_1 \l_2 & e^{-\t} \biggr) A_-^\dg(-\t) \cr
+ \wb e^{i\s} e^{-\t/2} 
\biggl( i M &\cosh\t  - {1\over 4}(\l_1^2 + \l_2^2) - 
{i\over 2} \l_1\l_2 e^\t \biggr) A_+^\dg(-\t). & \zps{b}\cr}$$
(We did not explicitly write down
the boundary operator $B$.)
Solving for the reflection S-matrix according to (\RSmseqZ) we find
$$ R_+^+(\t) = R_-^-(\t) = 
- { { \left( e^{\r(\t)} + e^{-\r(-\t)} \right) }\over
2\cosh(\t - \g(\t))} \eqno\zpa{a} $$
$$ R_+^-(\t) = i e^{- i\s} {1\over 2\cosh(\t - \g(\t))}
\left( e^{\t-\g(\t)+\r(\t)} - e^{-\t+\g(\t)-\r(-\t)}\right), \eqno\zps{b} $$
$$ R_-^+(\t) = i e^{+ i\s} {1\over 2\cosh(\t - \g(\t))}
\left( e^{\t-\g(\t)+\r(\t)} - e^{-\t+\g(\t)-\r(-\t)}\right), \eqno\zps{c} $$
\edef\RMsZ{\zen}%
where
$$ e^{\g(\t)} = {{\cosh\t - {i\over 4M}(\l_1^2 + \l_2^2) 
- {1\over 2M}{\l_1\l_2} e^\t}\over
{\cosh\t + {i\over 4M}(\l_1^2 + \l_2^2) - {1\over 2M}{\l_1\l_2} e^{-\t}} },
\quad e^{\r(\t)} = {{\cosh\t - {i\over 4M}(\l_1^2 + \l_2^2) 
- {1\over 2M}{\l_1\l_2} e^{-\t}}\over
{\cosh\t + {i\over 4M}(\l_1^2 + \l_2^2) - {1\over 2M}{\l_1\l_2} e^{-\t}} }.
\eqno\zp $$
Making use of the following relations
$$ e^{\r(\t)} e^{\r(-\t)} = e^{\g(\t)} e^{\g(-\t)} \eqno\zpa{a} $$
$$ e^{\g(i\pi/2 \pm \t)} = e^{-\g(i\pi/2\mp\t)},\quad
e^{\r(i\pi/2 \pm \t)} = e^{-\r(-(i\pi/2\mp\t))},\eqno\zps{b} $$
one can check that $R$ satisfies the unitarity and crossing relations 
[\Ghoshal]
$$ R_a^c(\t) R_c^b(-\t) = \delta_a^b,\quad
R_{\overline a}^b \left( {i\pi\over 2} - \t \right) =
-R_{\overline b}^a \left( {i\pi\over 2} + \t \right),\eqno\zp $$
where ${\overline a} = -a$.

For either $\l_1=0$ or $\l_2 =0$ we have
$$ e^{\r(\t)} = e^{\g(\t)} = e^{\r(-\t)} = e^{\g(-\t)}\qquad
{\rm if\ }\l_1=0{\ \rm or\ }\l_2=0. \eqno\zp $$
\edef\rgMKZ{\zen}%
The off-diagonal elements $R_+^-$ and $R_-^+$ differ only by a phase.
We will choose $\s$ such that $R_+^- = R_-^+$.  Thus from now on we take 
either $\s = 0$ or $\s=\pi$.
A specific choice will be made below.
Both cases can be obtained by applying a $U(1)$ transformation to the
fermion operators [\Ameduri].  For notational 
convenience define the following
$$ P(\t) \equiv R_+^+(\t) = R_-^-(\t),\quad Q(\t) \equiv R_-^+(\t) =
R_+^-(\t). \eqno\zp $$

The free fermion reflection amplitudes (\RMsZ) look quite 
complicated.  However by making a change of basis, their structure becomes
apparent.  The complex fermions can be expressed in terms of real fermions
$\p {1,2}$ and $\pb {1,2}$ 
$$ \p \pm = \fa(\p 1 \pm i\p 2 ),\quad \pb \pm = \fa(\pb 1 \pm i\pb 2 ).
\eqno\zp $$
In terms of these real fermions the action (\actionZ) becomes
$$\eqalignno{ S = & \, \, i\zit\zix
\sum_{i=1,2} \bigl[\p i(\d t - \d x)\p i + \pb i(\d t + \d x)\pb i + 
2 M \p i \pb i\bigl] 
\, - \, i\zit e^{i\s}\bigl( \p 1 \pb 1 - \p 2 \pb 2 \bigr)&\cr
& + {i\over 2}\zit
\bigl( a_1 \d t a_1 + a_2 \d t a_2 \bigr) 
- i {\lt_1 \over 2}\zit (\p 1 + e^{i\s} \pb 1) a_1 
- i {\lt_2 \over 2}\zit (\p 2 - e^{i\s} \pb 2) a_2, &\zp \cr} $$
\edef\IactZ{\zen}%
where
$$ \lt_1 = \l_1 e^{i\s} + \l_2,\quad \lt_2 = \l_1 e^{i\s} - \l_2,\eqno\zp$$
and we have defined the operators (not to be confused with the $a_\pm$)
$$ a_1(t) =  -{i\over\sqrt{2}}(\sp(t) - \sm(t)),\quad 
a_2(t) = \fa(\sp(t) + \sm(t)), \eqno\zp $$
satisfying
$$ \{a_1(t),a_2(t)\} = 0,\quad a_1(t)^2 = a_2(t)^2 = \fb. \eqno\zp$$
The action (\IactZ) is simply that of two decoupled Ising models with boundary
magnetic fields.  The reflection S-matrix for the Ising model 
was calculated in [\Ghoshal].  
Note that the boundary terms are
different for $(\p 1,\pb 1)$ and $(\p 2,\pb 2)$, with different couplings
in general.  The bulk and boundary equations of motion are 
$$(\d t - \d x) \p i + M \pb i = 0,\quad(\d t + \d x)\pb i - M \p i = 0
\quad (i=1,2)
\eqno\zp$$
\edef\IblkeomZ{\zen}%
$$ -\d t (\p 1 - e^{i\s} \pb 1) ={\lt_1^2 \over 4}(\p 1 + e^{i\s}\pb 1),\quad
-\d t(\p 2 + e^{i\s} \pb 2) = {\lt_2^2 \over 4}(\p 2 - e^{i\s} \pb 2),
\ \ ({\rm at\ }x=0), \eqno\zp$$
\edef\IbdyeomZ{\zen}%
which are just the Re and Im parts of (\FblkeomZ) and (\FbdyeomZ).
Mode expansions satisfying (\IblkeomZ) are given by
$$ \p i = \fmn\int_{-\infty}^\infty d\t\, 
e^{-\t/2}\bigl( \wb A_i(\t) e^{-i\vk\cdot \vx}
+\w A_i^\dg(\t) e^{i\vk\cdot \vx} \bigr) \eqno\zpa{a}$$
$$ \pb i = \fmn\int_{-\infty}^\infty d\t\, 
e^{\t/2}\bigl( \w A_i(\t) e^{-i\vk\cdot \vx}
+\wb A_i^\dg(\t) e^{i\vk\cdot \vx} \bigr), \eqno\zps{b}$$
\edef\ImeZ{\zen}%
where $A_{1,2}^\dg\ (A_{1,2})$ are the real fermion creation (annihilation)
operators satisfying (\AcomZ).  The complex fermion 
operators can be written as
$$ A_\pm = \fa (A_1 \mp i A_2),\quad 
A_\pm^\dg = \fa (A_1^\dg \pm i A_2^\dg). \eqno\zp$$
\edef\FopZ{\zen}%
Substituting (\ImeZ) into (\IbdyeomZ) we can obtain the 
reflection S-matrices for 
the real fermions and hence from (\FopZ) also for the complex fermions 
(solitons).  We find 
$$ A_{1,2}^\dg(\t) B = R_{1,2}(\t) A_{1,2}^\dg(-\t)B \eqno\zp$$
$$ R_{1,2}(\t) = {\widetilde R}_{1,2}(\t) f^{\rm CDD}_{1,2}(\t) \eqno\zp$$
\edef\RSmZ{\zen}%
$$ {\widetilde R}_{1,2}(\t) = 
i \tanh\bl i{\pi\over 4} \mp e^{i\s}{\t\over 2}\br \eqno\zp$$
$$ f^{\rm CDD}_{1,2}(\t) = - \left( 
{ {i\sinh\t + (\Delta_{1,2} \mp e^{i\s})}
\over{i\sinh\t -(\Delta_{1,2} \mp e^{i\s})} }
\right),\quad \Delta_{1,2} = { \lt_{1,2}^2\over 4 M}. \eqno\zp$$

The factor $\tilde{R}$ corresponds to a fixed boundary
condition for one copy of Ising fermion and a free boundary
condition for the other copy. 
In the absence of any interaction, the boundary equations are $(\s=0)$
$$ \p 1 - \pb 1 = 0,\quad \p 2 + \pb2 = 0, \eqno\zp$$
\edef\zcbciZ{\zen}%
corresponding to a free boundary for $(\p 1,\pb 1)$ and a fixed
boundary for $(\p 2,\pb 2)$.  
For infinite coupling $(|\lt_{1,2}|\rightarrow \infty)$ the boundary 
conditions are
$$ \p 1 + \pb 1 = 0, \quad \p 2 - \pb 2 = 0, \eqno\zp $$
\edef\icbciZ{\zen}%
that is a fixed boundary for $(\p 1, \pb 1)$ and a free boundary for
$(\p 2,\pb 2)$.   
(For $\s=\pi$ (\zcbciZ) and (\icbciZ) are interchanged.) 
So we have a flow from free to fixed (fixed to free) for $\p 1$ and $\pb 1$
($\p 2$ and $\pb 2$) as the interaction varies from zero to infinity.  In
particular, the flow is different for the two Ising models.  When expressed
in terms of the complex fermions, (\zcbciZ) and (\icbciZ) read
$$ \p + - \pb - =0,\quad \p - - \pb + = 0 \quad \quad (\lt_{1,2}=0)\eqno\zp $$
$$ \p + + \pb - =0,\quad \p - + \pb - = 0 \quad\quad (|\lt_{1,2}|=\infty).
\eqno\zp $$
Both equations correspond to free boundary conditions for the Dirac fermions,
and apart from a phase, give the same reflection amplitudes.  Thus there is no
flow between free and fixed boundary conditions for the solitons.  In contrast,
the MBSG model has a flow between free and fixed boundary conditions as the
coupling $\l$ goes from $0$ to $\infty$.

The soliton reflection amplitudes are given by
$$ P(\t)  = {1\over 2} (R_1(\t) + R_2(\t)),\quad
Q(\t) = {1\over 2} (R_1(\t) - R_2(\t)). \eqno\zp$$
\edef\RSmISZ{\zen}%
One can check that these are the same as (\RMsZ).  In particular we 
have 
$$ R_1(\t,\s=\pi) = R_2(\t,\s=0),\quad R_2(\t,\s=\pi) = R_1(\t,\s=0),
\eqno\zp $$
implying, as expected from (\RMsZ), that $P(\t)$ is independent of the 
choice for $\s$, whereas $Q(\t)$ changes by an overall sign.  

\zssecspc
\noindent
{\zssec \zsectwo.1. Free Fermion Reflection S-matrix for the MK model} \par
\nobreak
\zstsecspc
We now specialize to the MK model by taking
$$ \l_1 = \l,\quad \l_2 = 0.\eqno\zp $$
Using (\rgMKZ) we get for the free fermion reflection S-matrix
$$ P(\t) = - { \cosh \g(\t) \over \cosh(\t - \g(\t)) },\quad
 Q(\t) = i e^{i\s} {\sinh \t \over \cosh(\t - \g(\t)) }, \eqno\zp $$
\edef\MKRSmZ{\zen}%
with
$$ e^{\g(\t)} = {\cosh\t - i\Delta^{MK} \over \cosh\t + i \Delta^{MK} },
\quad \dd^{MK} = {\l^2\over 4 M},\eqno\zp $$ 
recovering the results of [\Leclair].
{}From (\RSmISZ) alternative expressions are
$$  P(\t) = - {i\over 2} \tanh \bl i {\pi\over 4} - {\t\over 2} \br
\bl { i\sinh\t + (\Delta^{MK}- 1)\over i\sinh\t -(\Delta^{MK}- 1) } \br
- {i\over 2} \tanh\bl i{\pi\over 4} + {\t\over 2}\br
\bl { i\sinh\t + (\Delta^{MK}+ 1)\over i\sinh\t -(\Delta^{MK}+ 1) } \br 
\eqno\zpa{a} $$
$$  Q(\t) = - e^{i\s}{i\over 2} \tanh \bl i {\pi\over 4} - {\t\over 2} \br
\bl { i\sinh\t + (\Delta^{MK}- 1)\over i\sinh\t -(\Delta^{MK}- 1) } \br
+ e^{i\s} {i\over 2} \tanh\bl i{\pi\over 4} + {\t\over 2}\br
\bl { i\sinh\t + (\Delta^{MK}+ 1)\over i\sinh\t -(\Delta^{MK}+ 1) } \br. 
\eqno\zps{b} $$
\edef\MKRSmIZ{\zen}%
Note that the two CDD factors in (\MKRSmIZ) are different.  The zero and
infinite coupling limits are (the upper (lower) sign corresponds to
$\l = 0\ (\infty)$)
$$ P(\t) = - \sech\t,\quad Q(\t) = \pm i e^{i\s} \tanh\t,\eqno\zp  $$
\edef\MKRinfZ{\zen}%
which are the same (apart from phases) as the MBSG reflection amplitudes
with free boundary conditions ($\l = 0$ in (\MBSGhamZ)).  
These results confirm that in the soliton basis there is no flow
between free and fixed boundary conditions for zero and infinite coupling.
This can also be seen from the boundary equations (\FbdyeomZ) with $\l_2=0$.

To check if (\MKRSmIZ) agrees with the Kondo model 
we compute the massless limit.
The massless limit for right-movers is obtained by letting 
$\t\rightarrow \t +\a$, and taking $\a\rightarrow \infty$, $M\rightarrow 0$
while keeping $M_0\equiv M e^\a/2$ held fixed.  This gives us the 
massless dispersion relation
$  E=p=M_0 e^\t$.
Taking the massless limit we find
$$ {\widetilde R}_{1,2} \rightarrow \mp i e^{i\s},\quad
f^{\rm CDD}_{1,2} \rightarrow 
-\tanh\left( {\t-\t_B^{MK}\over 2}-{i \pi\over 4}\right) \eqno\zp$$
$$ P(\t) = 0,\quad 
Q(\t) = i e^{i\s} \tanh\left( {\t-\t_B^{MK}\over 2}-{i \pi\over 4}\right),
\eqno\zp $$
\edef\MKRSlmlZ{\zen}%
where we have defined 
$$ M_0 \exp(\t_B^{MK}) = {\l^2\over 4}. \eqno\zp $$
The massless reflection S-matrix is almost entirely
due to the CDD factors $f_{1,2}^{\rm CDD}$.
Apart from a phase, this is the spin 1/2 reflection S-matrix
for the Kondo model [\Fendley].  The Kondo 
Hamiltonian (\KhamZ) is also recovered 
in the massless limit.  Thus the MK model reproduces the Kondo 
results at the free fermion point.

\zssecspc
\noindent
{\zssec \zsectwo.2. Free Fermion Reflection S-Matrix for the mMK model}\par
\nobreak
\zstsecspc
The mMK model is obtained by setting
$$ \l_1 = \sqrt{2 M} e^{\ph_0},\quad \l_2 = \sqrt{2 M} e^{-\ph_0}.
\eqno\zp$$
\edef\mMKcZ{\zen}%
For these values
$$ e^{\g(\t)} = -1,\quad e^{\r(\t)} = 
{ {i\sinh\t + {1\over 4 M}(\l_1^2 + \l_2^2)}\over
 {i\sinh\t - {1\over 4 M}(\l_1^2 + \l_2^2)} }, \eqno\zp $$
giving the reflection S-matrix
$$ P(\t) = \sech\t f^{mMK}(\t), \quad
 Q(\t) = i e^{i\s} \tanh\t f^{mMK}(\t), \eqno\zp $$
\edef\mMKRSmZ{\zen}%
where $f^{mMK}(\t)$ is the CDD factor
$$ f^{mMK}(\t) = e^{\r(\t)} = \left( { i\sinh\t + \dd^{mMK} \over
i\sinh\t - \dd^{mMK}} \right),\quad \dd^{mMK} = {1\over 4 M}(\l_1^2 +\l_2^2)= 
\cosh2\ph_0. \eqno\zp $$
\edef\mMKCDDZ{\zen}%
The CDD factor $f^{mMK}(\t)$ is due to the factors $f^{\rm CDD}_{1,2}(\t)$ in
(\RSmZ).  With (\mMKcZ) we have $f_{1}^{\rm CDD}=f_2^{\rm CDD}$, and
$f^{mMK}$ is simply equal to $-f^{\rm CDD}_{1,2}$. 
The hyperbolic factors in (\mMKRSmZ)
come from the ${\widetilde R}_{1,2}(\t)$ linear combinations
$$ {1\over 2}( {\widetilde R}_{1} + {\widetilde R}_{2} ) = -\sech(\t),
\quad {1\over 2}( {\widetilde R}_{1} - {\widetilde R}_{2} ) =
-i e^{i\s} \tanh\t. \eqno\zp$$
We see that the reflection amplitudes are the free MBSG amplitudes
(at the free fermion point) multiplied by a CDD factor.  The dependence on the
boundary coupling is solely in the CDD factor.  These elements are simpler 
than those of the MK model (\MKRSmIZ) where the 
two CDD factors are different.
In contrast to the MK model, there is no zero coupling limit.  {}From (\mMKcZ)
it is clear that both $\l_1$ and $\l_2$ cannot vanish for a non-zero mass.  
We can take an infinite coupling limit by letting $\ph_0\rightarrow \infty$
or $\ph_0\rightarrow -\infty$, corresponding respectively to 
$(\l_1,\l_2) \rightarrow (\infty,0)$ and 
$(\l_1,\l_2) \rightarrow (0,\infty)$.  In either case we get
$$ P(\t) =  -\sech\t,\quad Q(\t) = - i e^{i\s} \tanh\t,\eqno\zp  $$
which are the same as (\MKRinfZ).

The massless limit is taken as before with one important difference.  Because
the couplings are mass dependent, we must also let 
$\ph_0\rightarrow\infty$ while keeping $\l_1$ fixed.  Of course $\l_2$
goes to zero.  
Explicitly for the massless limit of $f^{mMK}$ we have
$$ f^{mMK}(\t) = { iM\sinh\t + M\cosh 2\ph_0\over iM \sinh\t - M\cosh 2\ph_0}
\longrightarrow { iM_0 e^\t + M_0 e^{\t^{mMK}_B} \over 
iM_0 e^\t - M_0 e^{\t^{mMK}_B} } = 
\tanh\left( {\t-\t_B^{mMK}\over 2}-{i \pi\over 4}\right), \eqno\zp $$
where $\t^{mMK}_B$ is defined through
$$ M_0 \exp(\t_B^{mMK}) = {M\over 2}e^{2\ph_0} = {\l_1^2\over 4}.\eqno\zp $$
The resulting reflection amplitudes are 
identical to (\MKRSlmlZ)
$$ P(\t) = 0, \quad 
Q(\t) = i e^{i\s} \tanh\left( {\t-\t_B^{mMK}\over 2}-{i \pi\over 4}\right).
\eqno\zp $$
%
Again we recover the Kondo results.  Since $\l_2\rightarrow 0$, the mMK
Hamiltonian (\mMKhamZ) goes over to the Kondo Hamiltonian (\KhamZ).

\advsecnum

\zmsecspc
\noindent
{\zmsec \zsecthree. Reflection S-Matrices away from the Free Fermion point 
and Integrability}\par
\nobreak
\zmtsecspc
In an integrable  boundary field theory, 
knowledge of the particle spectrum (both the
bulk spectrum and the boundary states) can allow one to  determine
the reflection S-matrix, up to CDD factors, by imposing the constraints
of boundary factorizability (i.e. boundary Yang-Baxter (BYB) equation),
unitarity and crossing symmetry [\Ghoshal],[\Fring].  For 
both spin 1/2 massive Kondo
models, the bulk spectrum is the sine-Gordon spectrum.  Also as mentioned
above, we will not explicitly consider any boundary state structure 
as in (\RSmbbeqZ).  
Thus if the massive Kondo models are 
integrable their reflection S-matrices must satisfy the same constraints
as for the MBSG model. 

The general
solution to the boundary 
scattering constraints for a sine-Gordon bulk spectrum, 
at arbitrary
$\b$, was described in GZ [\Ghoshal].  
The GZ 
solution depends on two formal parameters.  The difference between the
MBSG, MK and mMK reflection amplitudes will then be in the relation of these 
formal parameters to the physical parameters (mass, couplings and $\b$).
If the results of section \zsectwo\  are extendible 
to arbitrary $\b$, then the free
fermion matrices should agree with those of GZ at $\b=\sqrt{4\pi}$.

Before making this comparison let us review GZ's solution.
Define
$$ \ll = {8\pi\over \b^2} - 1.\eqno\zp $$
 For arbitrary $\ll>0$, 
GZ's solution to the scattering equations is 
$$ P^\gz_+(\t) \equiv R_+^+(\t) =  \cos(\xi -i \ll \t) r(\t),\eqno\zpa{a} $$
$$ P^\gz_-(\t) \equiv R_-^-(\t) =  \cos(\xi +i \ll \t) r(\t),\eqno\zps{b} $$ 
$$ Q^\gz_+(\t) \equiv R_+^-(\t) = Q^\gz_-(\t) \equiv R_-^+(\t) \equiv
Q^\gz(\t) = -{k\over 2} \sin(2i\ll\t) r(\t).\eqno\zps{c} $$
\edef\GZmgZ{\zen}%
The function $r(\t)$ can be written as
$$ r(\t) = r_0(\t) r_1(\t). \eqno\zp $$
The factor $r_0(\t)$ is independent of the boundary couplings and ensures 
that the crossing symmetry constraint is satisfied.  The boundary
dependence is contained in $r_1(\t)$, which takes the form
$$ r_1(\t) = {1\over \cos\xi} \s(\e,\t)\s(i\vt,\t). \eqno\zp $$
Here $\s(x,\t)$ is a function not to be confused with 
the phase $\s$.
The variables $k,\xi,\e$ and $\vt$ are formal parameters with no 
$\t$ dependence and satisfy 
$$ \cos\e\cosh\vt = -{1\over k} \cos\xi,\quad
\cos^2\e + \cosh^2\vt = 1 + {1\over k^2}.\eqno\zp $$
Thus only two of $k,\xi,\e$ and $\vt$ are independent.
These parameters have to be related to the physical parameters, which is
often the hardest part.  The functions $r_0(\t)$ and $\s(x,\t)$ can be
expressed as infinite products of $\Gamma$-functions (see [\Ghoshal]).  
Making use of [\Grad]
$$   \ln \Gamma(z) = \int_0^{+\infty} {dt\over t}\, \left[
(z-1) e^{-t} - {e^{-t} - e^{-zt} \over 1 - e^{-t}} \right],
\quad {\rm Re}(z)>0, \eqno\zp $$
the infinite $\Gamma$ products can be shown to have integral representations,
giving for $r_0$ and $\s$
$$ \s(x,\t) = {\cos x\over \cosh( \lc \t + ix)}
\exp\left[ i\int_{-\infty}^{\infty}\, dy\,
\sin(2\lc\t y/\pi) {\sinh\left(\lc + {2\over \pi} x\right)y
\over 2 y\cosh\lc y \sinh y} \right] \eqno\zp $$
\edef\GZsZ{\zen}%
$$ r_0(\t) = 
\exp\left[ -i\int_{-\infty}^{\infty}\, dy\,
\sin(2\lc\t y/\pi)
{\sinh{3\over 2}\lc y \sinh\left( {\lc-1\over 2}\right)y
\over  y\sinh{y\over 2}\sinh2\lc y} \right]. \eqno\zp $$
\edef\GZrZ{\zen}%
We will take $\xi = n\pi$ and write for the diagonal amplitude
$$ P^\gz(\t) \equiv P^\gz_+(\t) = P^\gz_-(\t) =
\cos(\xi)\cos(i \lc \t) r(\t).\eqno\zp $$
At the free fermion point, $\lc=1$, the expressions for the amplitudes
simplify greatly 
$$ P^\gz(\t) = -{ (\cos\xi\cosh\t)/k \over
(-i\sinh \t + \cos\e)(-i\sinh \t+ \cosh\vt )} =
{\cos\e\cosh\vt \cosh\t \over
(-i\sinh \t + \cos\e)(-i\sinh \t+ \cosh\vt )} \eqno\zp $$
\edef\GZpffZ{\zen}%
$$ Q^\gz(\t) = { {i \sinh \t\cosh \t}\over
(-i\sinh \t + \cos\e)(-i\sinh\t+ \cosh\vt )}.\eqno\zp $$
\edef\GZqffZ{\zen}%
In particular, if 
$$ \cos\e = - \cosh\vt = \pm 1, \eqno\zp $$
\edef\etavtZ{\zen}%
or equivalently
$$ k = \cos\xi = \pm 1, \eqno\zp $$
\edef\kxiZ{\zen}%
we have (for either sign)
$$ P^\gz_{\rm free}(\t) =\sech\t,\quad 
Q^\gz_{\rm free}(\t) = -i\tanh\t, \eqno\zp $$
\edef\GZfZ{\zen}%
which are the MBSG amplitudes for free boundary conditions.

\zssecspc
\noindent
{\zssec \zsecthree.1. The MK Reflection S-Matrix 
at the Reflectionless points}\par
\nobreak
\zstsecspc
In this subsection we propose a  natural extension of the
MK reflection S-matrix away from the free fermion point. 
The MK reflection S-matrix (\MKRSmZ) at the free fermion point can be 
rewritten as
$$ P(\t) = {\cosh\g(\t) \cosh(\t) \over 
(-i\sinh\t + 1)(-i\sinh\t +\cosh\g(\t))} \bl{i\sinh +(\dd^{MK} - 1)\over
i\sinh - (\dd^{MK} - 1)}\br \eqno\zpa{a} $$
$$ Q(\t) = {-i e^{i\s} \sinh(\t) \cosh(\t) \over 
(-i\sinh\t + 1)(-i\sinh\t +\cosh\g(\t))} \bl {i\sinh +(\dd^{MK} - 1)\over
i\sinh - (\dd^{MK} - 1)} \br. \eqno\zps{b} $$
\edef\MKffZ{\zen}%
Comparing these with (\GZpffZ) and (\GZqffZ) we see that 
$$ P(\t) = P^\gz(\t,\e=0,\vt=\g(\t))\times f^{\rm CDD}(\t),\quad
Q(\t) =  -e^{i\s} Q^\gz(\t,\e=0,\vt=\g(\t))\times f^{\rm CDD}(\t), 
\eqno\zp $$
where
$$ f^{\rm CDD} = {i\sinh +(\dd^{MK} - 1)\over i\sinh - (\dd^{MK} - 1)}.
\eqno\zp $$
The phase difference between $P(\t)$ and $Q(\t)$ is irrelevant, since if
$(P,Q)$ satisfy the scattering equations then so do $(P,-Q)$.  We will set
$\s=\pi$ for the MK model.  Up to a CDD factor, we see that the MK reflection 
amplitudes agree with those of GZ provided we take $\e=0$ and more 
importantly, allow $\vt$ to have the $\t$ dependence
$ \vt = \g(\t)$.
However, such a $\t$ dependence is not allowed by 
the BYB constraint [\Ghoshal].
Unitarity and crossing symmetry alone do not rule out the possibility of 
a $\t$ dependence for the parameters $(\xi,k,\e,\vt)$.  If there is no
BYB constraint, as is the case at the free fermion point, and more generally
at the reflectionless points, then (\MKffZ) is a possible solution to the
scattering equations.  Away from the reflectionless points, the BYB equation
forces the scattering amplitudes to take the general form of GZ's solution
(\GZmgZ) (up to CDD factors), with the formal parameters $\t$ 
independent.  This means that (\MKffZ) cannot be extended to arbitrary values 
of $\lc$ in a manner consistent with (\GZmgZ).  Since GZ's solution 
is the unique solution to the boundary 
scattering equations for a sine-Gordon spectrum at arbitrary $\lc$, 
we conclude that assuming there are no boundary degrees of freedom, 
the  MK model is not integrable (i.e. there is no reflection
S-matrix consistent with BYB, unitarity and crossing symmetry) for generic
$\lc$.  Again we would like to emphasize that for higher spin $j\geq 1$, one
probably needs to solve a more general BYB equation with boundary spin states
and amplitudes $R_{a\a}^{b\b}(\t)$.  In which case a comparison with the
MBSG model cannot be made.

The reflectionless points occur at integer $\lc$, for which the BYB 
constraint is trivially satisfied.  The reason for this being that if
$\lc=n\geq 1$, the bulk scattering matrix takes on a simple form 
with only one independent scattering amplitude.  In general the
bulk sine-Gordon scattering matrix is 
$$ a(\t) \equiv S_{++}^{++}(\t) = S_{--}^{--}(\t) = \sin[\lc(\pi+i\t)]Z(\t)
\eqno\zpa{a} $$
$$ b(\t) \equiv S_{+-}^{+-}(\t) = S_{-+}^{-+}(\t) = -\sin(i\lc\t)Z(\t)
\eqno\zps{b} $$
$$ c(\t) \equiv S_{+-}^{-+}(\t) = S_{-+}^{+-}(\t) = \sin(\lc\pi)Z(\t),
\eqno\zps{c} $$
where the function $Z(\t)$ can be found in [\Zam].  For integer $\lc$ we
have
$$ b(\t) = a(\t)\ (\lc {\rm \ odd})\ \ {\rm or} \ \ b(\t) = -a(\t)\ 
(\lc {\rm \ even}), \quad {\rm and} \ \ c(\t) = 0. \eqno\zp $$
\edef\SGIbmZ{\zen}%
The expression for $a(\t)$ reduces to
$$ a(\t) = (-1)^\lc \prod_{n=1}^{\lc-1} f_{n/\lc}(\t), \eqno\zp $$
where
$$f_\alpha = {{\sinh \left({1\over2}[\t + i\pi \alpha] \right)}\over
{\sinh \left({1\over2}[\t - i\pi\alpha] \right)}}.\eqno\zp $$
\edef\falphaZ{\zen}%
One can check that for (\SGIbmZ) there is no constraint from the BYB equation. 
Since the off-diagonal bulk amplitude $c(\t)$ vanishes, 
these integer points are
referred to as the reflectionless points.  

With no BYB constraint, it
should be possible to extend (\MKffZ) to the reflectionless points.
In terms of $R_{1,2}$, the unitarity and crossing relations for 
integer $\lc$ become

\smallskip
\noindent
Unitarity:
$$ R_1(\t) R_1(-\t) = 1,\quad R_2(\t) R_2(-\t) = 1.\eqno\zp$$
\edef\unZ{\zen}%

\noindent
Crossing:

\noindent
Odd $\lc$
$$ R_1(i\pi/2 - \t) = a(2\t)\, R_1(i\pi/2 + \t),\quad
R_2(i\pi/2 - \t) = a(2\t)\, R_2(i\pi/2 + \t)\eqno\zpa{a} $$

\noindent
Even $\lc$
$$ R_1(i\pi/2 - \t) = -a(2\t)\, R_2(i\pi/2 + \t),\quad
R_2(i\pi/2 - \t) = -a(2\t)\, R_1(i\pi/2 + \t).\eqno\zps{b} $$
\edef\crsZa{\zen}%

\smallskip
\noindent
Since $a(-\t)=1/a(\t)$, the two crossing equations for even $\lc$ are 
identical.

Now recall the free fermion reflection amplitudes in the Ising notation 
(\RSmISZ), where $(\s=\pi)$ 
$$ R_{1,2}(\t) = {\widetilde R}_{1,2}(\t) f^{\rm CDD}_{1,2}(\t) \eqno\zp$$
\edef\RamZ{\zen}%
$$ {\widetilde R}_{1,2}(\t) = 
i \tanh\bigl(i{\pi\over 4} \pm {\t\over 2}\bigr),\quad 
 f^{\rm CDD}_{1,2} = - \left( 
{ {i\sinh\t + (\Delta^{MK} \pm 1)}\over{i\sinh\t -(\Delta^{MK} \pm 1)} }
\right). \eqno\zp $$
\edef\RfacZ{\zen}%
Considering the structure of the reflection 
matrices (\RamZ) for $\lc=1$, we will 
look for solutions of (\unZ) and (\crsZa) of the following form
$$ R_1(\t) = {\tilde r}_1(\t) f_1^{\rm CDD}(\t),\quad
R_2(\t) = {\tilde r}_2(\t) f_2^{\rm CDD}(\t),\eqno\zp$$
where ${\tilde r}_{1,2}(\t)$ are independent of the boundary coupling $\l$.
The boundary dependence being contained in the CDD factors.
For odd $\lc$, the CDD
factors can be different for $R_1$ and $R_2$, as they are for $\lc=1$.  
However for even $\lc$, since $a(2\t)$ is independent of $\l$, 
the crossing relation implies that the CDD factors must be the same.  
Requiring the CDD factors to agree with (\RfacZ) for $\lc=1$ suggests we
take the general form 
$$ f^{\rm CDD}_{1,2} = 
- { {i\sinh\t + \gg_{1,2}(\lc,\l^2/G) }\over
  {i\sinh\t - \gg_{1,2}(\lc,\l^2/G) } }, \eqno\zp$$
where $\gg_{1,2}(\lc,\l^2/G)$ are functions of the dimensionless parameters
$\lc$ and $\l^2/G$, satisfying
$$ \gg_{1,2} \bl \lc=1,{\l^2\over G}\br = \dd^{MK} \pm 1 \eqno\zp $$
\edef\ggffcZ{\zen}%
$$ \gg_{1} \bl \lc=2n, {\l^2\over G}\br = \gg_2 \bl \lc =2n, {\l^2\over G}\br.
\eqno\zp $$
\edef\ggecZ{\zen}%
Note that (\ggffcZ) makes sense since at the free fermion point $M\propto G$.

The crossing relations that need to be satisfied can now be written as
$$ a(2\t)\,{\tilde r}_{1,2}(\t + i\pi/2){\tilde r}_{1,2}(\t - i\pi/2)= 1\ 
(\lc\ {\rm odd})
\eqno\zpa{a}$$ 
$$ -a(2\t)\,{\tilde r}_{1}(\t + i\pi/2){\tilde r}_{2}(\t - i\pi/2)= 1\ 
(\lc\ {\rm even})
\eqno\zps{b}$$
\edef\crsZb{\zen}%
To solve these equations we make use of the following expression for
$a(2\t)$
$$ a(2\t) = \prod_{n=1}^{\lc} f_{-n/2\lc}(\t - i\pi/2)
f_{-n/2\lc}(\t+i\pi/2),\eqno\zp $$
where the $f_\alpha$'s are given by (\falphaZ).  Since 
$$ f_{1/2}(\t) = {\widetilde R}_1(\t), \eqno\zp$$
we take ${\tilde r}_1(\t)$ to be
$$ {\tilde r}_1(\t) = \prod_{n=1}^{\lc} f_{n/2\lc}(\t). \eqno\zp $$
\edef\rooZ{\zen}%
This will satisfy the odd $\lc$ crossing relation for ${\tilde r}_1(\t)$.  
Similarly, knowing that
$$ -f_{1/2}(\t-i\pi) = {\widetilde R}_2(\t),\eqno\zp $$ 
we take for ${\tilde r}_2(\t)$ 
$$ {\tilde r}_2(\t) = -\prod_{n=1}^{\lc} f_{n/2\lc}(\t -i\pi). \eqno\zp $$
\edef\rtoZ{\zen}%
Using
$$ f_\alpha(\t + i2\pi m ) = f_{\a+ 2m}(\t) = f_\alpha(\t),\eqno\zp $$
\edef\periodpfZ{\zen}%
where $m$ is any integer, 
we see that (\rtoZ) satisfies the odd $\lc$ crossing relation for $\rt_2(\t)$.
Unitarity for ${\tilde r}_{1,2}$ is also easily checked with the
relations
$$ f_\alpha(\t)f_\alpha(-\t)=1,\quad f_\a(-\t) = f_{-\a}(\t). \eqno\zp$$
The solutions (\rooZ) and (\rtoZ) are not valid for even $\lc$.  These
expressions imply
$$ \rt_1(\t + i\pi/2) \rt_2(\t - i\pi/2) = 
-\prod_{n=1}^\lc f_{n/2\lc}^2(\t+i\pi/2),\eqno\zp$$
which does not satisfy the crossing relation (\crsZb b).  Possible solutions
for (\crsZb b) are
$$ \rt_1(\t) = -\rt_2(\t) = \prod_{n=1}^{\lc} f_{n/2\lc}(\t),\eqno\zp$$
and
$$ \rt_1(\t) = -\rt_2(\t) = \prod_{n=1}^{\lc} f_{n/2\lc}(\t - i\pi).
\eqno\zp $$
We can now combine the odd and even $\lc$ solutions into a single expression.
For any integer $\lc$ we have
$$ \rt_1(\t) = \prod_{n=1}^{\lc} f_{n/2\lc}\left(\t -i\pi \rho_1(\lc)\right)
\eqno\zp $$
$$ \rt_2(\t) = -\prod_{n=1}^{\lc} f_{n/2\lc}\left(\t -i\pi \rho_2(\l)\right)
\eqno\zp $$
where for odd $\lc$ 
$$ \rho_1(\lc=2n+1) = 0,\quad \rho_2(\lc =2n+1) = 1, \eqno\zp $$
and for even $\lc$ we can have two choices
$$ \rho_1(\lc=2n) = 0,\quad \rho_2(\lc = 2n) = 0, \eqno\zpa{a} $$
or
$$ \rho_1(\lc=2n) = 1,\quad \rho_2(\lc = 2n) = 1. \eqno\zps{b} $$
\edef\rhoevenZ{\zen}%
These expressions for $\rho_{1,2}(\lc)$ are mod 2 because of (\periodpfZ).
If it were not for the specific forms (\RfacZ) for $\lc=1$, we would also
have two choices for odd $\lc$. 
As an example, we can take for (\rhoevenZ a)
$$ \rho_1(\lc) = 0,\quad \rho_2(\lc) = \sin^2(\lc\pi/2).\eqno\zp $$

Putting everything together we have for any integer $\lc$
$$ R_{1,2}(\t) = \mp \bl{ {i\sinh\t + \gg_{1,2}(\lc,\l^2/G) }\over
  {i\sinh\t - \gg_{1,2}(\lc,\l^2/G) } } \br
\prod_{n=1}^{\lc} f_{n/2\lc}\left(\t -i\pi \rho_{1,2}(\lc)\right), 
\eqno\zp $$
giving for the soliton reflection S-matrix at the reflectionless points
$$ P(\t) = \fb\sum_{j=1,2}(-1)^j
\bl { {i\sinh\t + \gg_j(\lc,\l^2/G) }
\over {i\sinh\t - \gg_j(\lc,\l^2/G) } } \br
\prod_{n=1}^{\lc} f_{n/2\lc}\left(\t -i\pi \rho_{j}(\lc)\right) 
\eqno\zpa{a} $$
$$ Q(\t) = -\fb\sum_{j=1,2} 
\bl { {i\sinh\t + \gg_j(\lc,\l^2/G) }
\over {i\sinh\t - \gg_j(\lc,\l^2/G) } } \br
\prod_{n=1}^{\lc} f_{n/2\lc}\left(\t -i\pi \rho_{j}(\lc)\right). 
\eqno\zps{b} $$
\edef\MKrpZ{\zen}%
Note that for even $\lc$ 
$$ P(\t) = 0, \quad \lc\ {\rm even}.\eqno\zp $$
Further restrictions on $\gg_{1,2}(\lc,\l^2/G)$ and $\rho_{1,2}(\lc)$ can be 
obtained by analyzing the pole structure (see section \zsecfour), and in 
addition $\gg_{1,2}(\lc,\l^2/G)$ is also constrained by the massless limit.

The massless limit for arbitrary $\lc$ is obtained by taking
$G\rightarrow 0$, along with $\t\rightarrow \t+\a$ and $\a\rightarrow\infty$,
while keeping $G_0 = G^{(\lc+1)/2\lc} e^\a/2$ held fixed.  In general
$M\propto G^{(\lc + 1)/2\lc}$, thus $G^{(\lc + 1)/2\lc}$ acts as a mass,
with $G_0$ replacing $M_0$ away from the free fermion point.  This gives our
``massless limit prescription''.  In order for the massless limit to make 
sense and be of the form (\MKRSlmlZ), the limits 
$\lim_{G\rightarrow 0} G^{(\lc+1)/2\lc} \gg_{1,2}(\lc,\l^2/G)$ must be
well-defined and equal. 
Considering (\ggffcZ), a reasonable choice for $\gg_{1,2}$
is to take 
$$ \gg_{1,2}\bl \lc,{\l^2\over G} \br = 
c(\lc) {\bl {\l^2 \over G} \br}^{(\lc+1)/2\lc} +\tau_{1,2}(\lc), \eqno\zp$$
where $c(\lc)$ and $\tau_{1,2}(\lc)$ are functions chosen to satisfy (\ggffcZ)
and (\ggecZ).  The values of $\tau_{1,2}(\lc)$ are irrelevant in the massless
limit.  Following this procedure we find
$$ \rt_{1,2}(\t) \longrightarrow \pm i e^{i{\pi\over 4}(\lc-1)},\quad
f_{1,2}^{\rm CDD}(\t) \longrightarrow 
- \tanh\left( {{\t-\t_B^{MK}}\over 2} - {i\pi\over 4} \right) \eqno\zp $$
$$ P(\t) = 0,\quad Q(\t) = -i e^{i{\pi\over 4}(\lc-1)}
\tanh\left( {{\t-\t_B^{MK}}\over 2} - {i\pi\over 4} \right), \eqno\zp $$
where $\t_B^{MK}$ is $\lc$ dependent, defined to satisfy 
$$ G_0 \exp \t_B^{MK} = 
\lim_{G\rightarrow 0} G^{(\lc + 1)/2\lc} \gg_1 \bl \lc,{\l^2\over G}\br =
\lim_{G\rightarrow 0} G^{(\lc + 1)/2\lc} \gg_2 \bl \lc,{\l^2\over G}\br .
\eqno\zp $$
We recover the Kondo reflection S-matrix for all integer values of $\lc$.
Because $\rt_{1,2}(\t)$ is independent of the boundary, it only
contributes a phase.
This phase is actually necessary to satisfy the massless crossing relation.
In the massless unitarity relation the phase cancels since it appears
conjugated in $Q(-\t)$.  Massless unitarity and crossing relations
are discussed in [\Aludwig].

\zssecspc
\noindent
{\zssec \zsecthree.2. The mMK Reflection S-Matrix for general $\b$}\par
\nobreak
\zstsecspc
The mMK scattering amplitudes at the free fermion point
are easily related to GZ's general solution.  Comparing (\mMKRSmZ) and 
(\GZfZ) we find
$$ P(\t) = P^\gz_{\rm free}(\t) f^{mMK}(\t),\quad 
Q(\t) = Q^\gz_{\rm free}(\t) f^{mMK}(\t), \eqno\zp$$
\edef\mMKffZ{\zen}%
where we have set the phase $\s=\pi$.  These are just the  MBSG 
amplitudes with the free boundary condition 
multiplied by a CDD factor, with the formal parameters 
satisfying (\etavtZ) and (\kxiZ).   Without loss of generality, we will fix 
$\xi$ to be 0 for all $\lc$.  For real $\vt$, (\etavtZ) gives the values
$$ \e(\lc=1) = \pi \Longrightarrow k(\lc=1) = 1,\quad 
\vt(\lc=1) = 0.\eqno\zp $$
\edef\mMKetavtZ{\zen}%
Unlike the MK model, (\mMKffZ) and (\mMKetavtZ) are consistent with the 
generic result (\GZmgZ), thus allowing for an extension to arbitrary $\lc$.
Taking into account the structure of the 
free fermion amplitudes (\mMKffZ), we
propose the following soliton reflection S-matrix for the mMK model
$$ P(\t) = P_{\rm free}(\t) f^{mMK}(\t),\quad
Q(\t) = Q_{\rm free}(\t) f^{mMK}(\t),\eqno\zp $$
\edef\mMKgenZ{\zen}%
where $P_{\rm free}(\t)$ and $Q_{\rm free}(\t)$ are the MBSG
amplitudes for free boundary conditions 
$$ P_{\rm free}(\t) =\cosh(\lc \t ) r_0(\t)
\s\bl {4\pi^2\over \b^2},\t\br \s(0,\t)\eqno\zpa{a} $$
$$ Q_{\rm free}(\t) =-i\sinh(\lc \t) 
{\cosh(\lc\t)\over \sin(\lc\pi/2)} r_0(\t)
\s\bl {4\pi^2\over \b^2},\t\br \s(0,\t),\eqno\zps{b} $$
\edef\MBSGfmZ{\zen}%
and 
$$ f^{mMK} = {i\sinh\t + \gg(\lc,\ph_0)\over i\sinh\t - \gg(\lc,\ph_0) },
\eqno\zp $$
\edef\mmkggfZ{\zen}%
with $\gg(\lc,\ph_0)$ some function satisfying
$$ \gg(\lc=1,\ph_0) = \cosh 2\ph_0. \eqno\zp $$
The CDD factor (\mmkggfZ) is a generalization of (\mMKCDDZ) to arbitrary
values of $\lc$.
In (\MBSGfmZ) the parameter $\e$ has been given a $\lc$ dependence
$$ \e(\lc) = {\pi \over 2}(\lc + 1) = {4 \pi^2 \over \b^2}
\Longrightarrow k(\lc) = {1\over \sin(\lc\pi/2) },\eqno\zp $$
while $\vt(\lc)=\vt(\lc=1)=0$.  This serves to produce the correct pole
structure, namely a simple pole at $\t = i\pi/2$ for all $\lc$ [\Ghoshal]. 
It appears that $Q_{\rm free}$ is singular for even $\lc$ because of the 
$\sin(\lc\pi/2)$ factor.  But an identical factor appears in the numerator
of $\s(4\pi^2/\b^2,\t)$ (\GZsZ), making (\MBSGfmZ b) well-defined 
for all $\lc$.  
Since at the free fermion point $\l = \sqrt{2 M} \propto \sqrt{G}$, and $G$
is the only available bulk parameter, we must have $\l^2 \propto G$ for 
arbitrary $\lc$, thus explaining the relation (\mMKhamZ c).  

The main feature of (\mMKgenZ) is that
the boundary dependence occurs only in the CDD factor, just as for MK model.
The minimal (non-CDD) parts $P_{\rm free}$ and $Q_{\rm free}$ 
are independent of $\ph_0$.
In contrast, the boundary dependence for the MBSG model is in general
more complicated, with the formal parameters $\e$ and $\vt$ (and hence the 
factor $r_1(\t)$) being functions of the physical parameters.  
With the boundary dependence isolated, the massless limit of the minimal
factors follows by taking $\t\rightarrow\infty$.  
The exact dependence of $\e$ and $\vt$ on $\lc$
is not important in the massless limit as long as they are independent of the
boundary coupling. The massless limit for the CDD factor is taken in the
same manner as for the MK model, with the difference that $\ph_0$ also has
to be appropriately scaled such that $\sqrt{G} e^{\ph_0}$ is held fixed.
Holding $\sqrt{G} e^{\ph_0}$ fixed amounts to keeping the coupling $\l_1$
constant as $G\rightarrow 0$. The massless limit restricts $\gg(\lc,\ph_0)$ to
the extent that 
$\lim_{G\rightarrow 0,\ph_0\rightarrow\infty} 
G^{(\lc+1)/2\lc} \gg(\lc,\ph_0)$,
with $\sqrt{G} e^{\ph_0}$ fixed, must be well-defined.
Making use of the integral representations (\GZsZ) and
(\GZrZ) we find for the functions $r_0(\t)$ and $\s(x,\t)$
$$ \lim_{\t\rightarrow \infty} r_0(\t) = 1,\quad
\lim_{\t\rightarrow \infty} \s(x,\t) = 
2\cos(x) e^{-ix} \lim_{\t\rightarrow \infty} e^{-\lc\t},\eqno\zp $$
which gives for the massless amplitudes
$$ P(\t) = 0, \quad Q(\t) = i e^{-i{\pi\over 2}(\lc + 1)}
\tanh\left( {{\t-\t_B^{mMK}}\over 2} - {i\pi\over 4} \right), \eqno\zp $$
where $\t^{mMK}_B$ as a function of $\lc$ is defined through
$$ G_0 \exp \t_B^{mMK} = \lim_{G\rightarrow 0,\ph_0\rightarrow\infty}
G^{(\lc+1)/2\lc} \gg(\lc,\ph_0) {\Bigm |}_{\sqrt{G}e^{\ph_0}\ {\rm fixed}}.
\eqno\zp $$
So up to some phase we obtain the correct Kondo results for all $\lc$.
This suggests that the integrable massive generalization of Kondo is given 
by the mMK model (\mMKhamZ).  We again point out that
the situation is different here as compared with the
MBSG model.  For MBSG, $\e$ and $\vt$ are boundary
scale dependent.  To get the massless limit, these parameters also have
to be rescaled such that $g(\t,\e,\vt)\rightarrow \t -\t_B$ remains finite,
where $g$ is some function giving 
the massless limit prescription [\Warner].  
In our case the prescription (for the minimal part) is very simple,
let $\t\rightarrow\infty$. 

\advsecnum

\zmsecspc
\noindent
{\zmsec \zsecfour. Boundary Bound States}\par
\nobreak
\zmtsecspc
Important information on the spectrum of a theory can be obtained from the
poles of the reflection amplitudes.  Poles in the reflection S-matrix
$R$ can be of two types.  If there are bulk bound states interacting with the 
boundary, then poles will be found at $\t= i\pi/2$ and $\t = i\pi/2-\t_b$,
where $\t_b$ is the bound state pole in the bulk S-matrix.   These bulk states
will give zero-momentum one-particle contributions to the boundary state
$\rb B$.  Secondly, poles associated with boundary bound states can
also occur.  These are excitations of the boundary ground state
appearing in the region $0\leq \t \leq i\pi/2$. If the 
pole occurs at
$ \t = i\pi/2$, then there will be a degeneracy in the ground state and
again a zero-momentum state will be found in $\rb B$.  At the free fermion
point we only expect boundary bound states since there are no bulk bound 
states (the bulk S-matrix is $-1$).

\zssecspc
\noindent
{\zssec \zsecfour.1. Boundary bound states for the MK model}\par
\nobreak
\zstsecspc
At the free fermion point we can map the MK model to the two Ising
models, called Ising1 for $(\p 1,\pb 1)$ and Ising2 for $(\p 2,\pb 2)$, 
to study the poles.  These Ising models differ in their boundary
conditions at zero and infinite coupling.
The poles of the MK reflection S-matrix occur at
$$ \t_1 = i {\pi\over 2},\quad {\rm and}\quad \sin v_2 = 1- {\l^2\over 4M}
\ \ (\t_2=i v_2), \eqno\zp $$
where $\t_1$ is due to Ising1 and $\t_2=iv_2$, which 
only exists in the physical region if 
$\l^2 < 4M$, is due to Ising2.  As in [\Ghoshal] these have the following 
interpretation.  Each Ising model has two degenerate boundary ground states
${\rb {0,\pm}}_B^i \ \{i=1,2\}$, giving two different expectation values
for the spin field $\s(x)$, ${\langle \s(x)\rangle}_\pm \propto \pm 1$.
For the Ising2 piece, the degeneracy is broken by the boundary magnetic
field
$$   {\rb {0,\pm}}_B^2 \longrightarrow {\rb 0}_B^2,{\rb 1}_B^2,\eqno\zp $$
and the pole at $\t_2$ corresponds to the boundary bound state ${\rb 1}_B^2$
with energy
$$ E^{(2)}_1 - E^{(2)}_0 = M \cos v_2. \eqno\zp $$
In the Ising1 case the degeneracy persists for a non-zero boundary
interaction $(\l \not= 0)$.  Both states have the same 
energy
$$ E^{(1)}_+ = E^{(1)}_-. \eqno\zp $$
Thus for $\l^2< 4M$ we have two sets of degenerate states 
$$ \{ \rb{0,+}_B^1 \rb{0}_B^2, \rb{0,-}_B^1 \rb{0}_B^2 \}\quad {\rm and}\quad
\{ \rb{0,+}_B^1 \rb{1}_B^2, \rb{0,-}_B^1\rb{1}_B^2 \},\eqno\zp $$
\edef\BstatesMKZ{\zen}%
associated with the poles at $\t_1$ and $\t_2$ respectively.  
This  suggests  that one should actually regard the  
boundary as having structure
$B_\alpha$, where $\alpha = \pm$ corresponds to the above
degenerate states.    Because of the
pole at $i\pi/2$, there will be  a zero-momentum contribution to $\rb B$ of
the form $A_1^\dg(0) = \fa(A_+^\dg(0) +  A_-^\dg(0)$.  If $\l^2> 4M$,
the pole at $\t_2$ leaves the physical strip and we are left with the
degenerate ground states.

It is not clear what this pole structure implies for the states of $\phi$ and 
the spin operators $S_\pm$.  We make the following remark.  For a spin 1/2
impurity along with the screening effect, the boundary is described by
the Hilbert space
$$ \fb \otimes \fb = 0 \oplus 1.\eqno\zp $$
\edef\BhilbertZ{\zen}%
For an infinite interaction $(\l=\infty)$ the boundary would exist in the
$\rb{j,m} = \rb{0,0}$ state.  With a finite interaction $(\l^2<4M)$, 
the space (\BhilbertZ) decomposes into two sets of degenerate states,
which are some linear combinations of $\rb{0,0}$, $\rb{1,0}$ and 
$\rb{1,\pm 1}$.  These have to be related to the Ising states.

Away from the free fermion point the bulk S-matrix has soliton bound states. 
These are the charge neutral breathers with masses
$$ m_n = 2 M \sin\bl{n\pi\over 2 \lc}\br,\quad n=1,2,\ldots <\lc.\eqno\zp $$
The bound states should give rise to poles in the reflection amplitudes.
For odd $\lc>1$, the poles in $P(\t)$ (\MKrpZ a) at
$$ \t = i {n\pi\over 2 \lc},\quad n=1,2\ldots<\lc,\eqno\zp $$
\edef\brpZ{\zen}%
are associated with the breathers.  These poles also occur in $Q(\t)$, 
implying that the boundary should carry a $(+,-)$ ``topological charge" since 
the breathers are charge neutral.  
This would agree with a description of the boundary in terms of $B_\pm$, with
the degenerate states (\BstatesMKZ) having different charges.
To account for the breathers at even
$\lc>1$ we must take (\rhoevenZ a) for $\r_i(\lc)$, which will reproduce the
poles (\brpZ).  The diagonal amplitude $P(\t)$ vanishes for even $\lc$.  This 
would make sense if at even $\lc$ the boundary charge is not conserved, leading
to a change in the soliton charge.

With the choice (\rhoevenZ a), 
the pole at $\t_1 = i\pi/2$ is present for all integer $\lc$ 
and the ground state is twofold degenerate.  
The existence of a boundary bound state for $\lc>1$ depends on 
$\gg_{1,2}(\lc,\l^2/G)$.  If we have 
$-1<\gg_{1,2}(\lc,\l^2/G)<0$, then a pole will exist
in the physical strip at 
$$ \t_2 = i v_2, \quad \sin v_2 = - \gg_{1,2}(\lc,\l^2/G), \eqno\zp $$
leading to a bound state.  However for $\lc>1$, a map to the
Ising picture does not exist and the boundary states cannot be explained
as in (\BstatesMKZ).
For odd $\lc$ we can have $-1<\gg_{1,2}(\lc=2n+1,\l^2/G)<0$ and
$\gg_1 \neq \gg_2$, which will lead to two distinct bound states.

\zssecspc
\noindent
{\zssec \zsecfour.2. Boundary bound states for the mMK model}\par
\nobreak
\zstsecspc
The pole structure for the mMK model is identical to that of the MBSG
model with free boundary conditions, apart from any poles due to the
CDD factor.
We will only comment
on the boundary bound states.  At the free fermion point the mMK model can be 
written as a sum of two Ising models, Ising1 and Ising2, with
couplings $\lt_1$ and $\lt_2$. Consider for the moment the more
general system (\IactZ).  The first Ising copy leads to the pole
at $\t_1 = i\pi/2$ as for the MK model.  The second copy contains the
CDD factor 
$$ f^{\rm CDD}_{2} = - \left( 
{ {i\sinh\t + (\Delta_2 - 1)}\over{i\sinh\t -(\Delta_2 - 1)} }
\right), \eqno\zp $$
which gives a boundary bound state provided the ``magnetic field" is not
too strong 
$$ \lt_2^2 < 4M. \eqno\zp $$
In terms of $\l_1$ and $\l_2$ ($\s=\pi$)
$$ \lt_2 ^2  = \l_1^2 + \l_2^2 + 2\l_2\l_2. \eqno\zp $$
Now specialize to the mMK system with
$$ \l_{1,2} = \sqrt{2M} e^{\pm \ph_0}, \eqno\zp $$
giving
$$ \lt_2^2 = 4M(\cosh 2\ph_0 + 1) > 4M. \eqno\zp $$
Thus there is no boundary bound state because the coupling $|\lt_2|$
is too large for all values of $\ph_0$.  The poles for the CDD factor
$f^{mMK}(\t) $
$$ f^{mMK}(\t) = -f_2^{\rm CDD}(\t) =  \left( 
{ {i\sinh\t + \cosh 2\ph_0 }\over{i\sinh\t - \cosh 2\ph_0} }
\right), \eqno\zp $$
occur outside the physical strip at
$$ \t_0 = \pm 2\ph_0 - i{\pi\over 2}. \eqno\zp $$
So the boundary states at $\lc=1$ consist of two ground states
$$ \{ \rb{0,+}_B^1 \rb{0}_B^2, \rb{0,-}_B^1 \rb{0}_B^2 \},\eqno\zp $$
corresponding to the pole at $i\pi/2$.

By choosing $\e = {\pi\over 2}(\lc+1)$, we maintain the pole at $i\pi/2$ for
all $\lc$, though its interpretation in terms of the Ising states is lost
for $\lc>1$. Away from the free fermion point $f^{mMK}(\t)$ has a pole at
$$ \sinh \t_2 = -i\gg(\lc,\ph_0). \eqno\zp $$
If $-1<\gg(\lc,\ph_0)<0$, this pole will lie in the physical strip implying a
bound state.

\zmsecspc
\noindent
{\zmsec \zsecfive. Conclusions}\par
\nobreak
\zmtsecspc
We have presented two massive versions of the anisotropic spin 1/2
Kondo model and discussed their integrability.  Both models 
correspond to a sine-Gordon theory in the bulk, 
but differ slightly in their 
boundary interactions.
Our arguments for integrability were  based on the assumption
that there are no explicit boundary degrees of freedom in the scattering
description, which is supported by the screening at
the free fermion point,  on constraints from  the
boundary Yang-Baxter equation,  and 
comparison with the massive boundary sine-Gordon  model at the free
fermion point.
We argued that the 
most natural massive extension of the Kondo model (MK) is not integrable
except perhaps at the reflectionless points, and  we proposed a reflection
S-matrix for these points.  
The modified massive Kondo model (mMK), on the other hand,  we argued is
integrable for all $\beta$  
and  proposed an exact reflection
S-matrix, which is a  CDD factor times  the massive boundary
sine-Gordon solution corresponding to a  free boundary condition. 

Since we did not prove the integrability of our models by studying integrals
of motion, our proposed S-matrices and the claim that the modified massive
Kondo model is integrable remain conjectures.  However, the fact that we were
able to construct an exact S-matrix, starting with an explicit calculation
at the free fermion point and consistently (i.e. agreeing with the scattering
constraints) extending the result to arbitrary $\beta$, supports the
integrability claim.  A more concrete analysis would involve trying to 
construct quantum conserved charges for arbitrary $\beta$.  Compared to
the massive boundary sine-Gordon model
and the (massless) Kondo model, such an analysis for the massive Kondo models
is more complicated due to the presence of both (i) interactions in the bulk
and at the boundary and (ii) spin operators $S_\pm$ at the boundary.
In relation to this, it would be interesting to develop conformal perturbation
theory to study integrability in the presence of boundary operators such
as $S_\pm$.  Lastly, a classical study of the massive models can also be
attempted.  Classical integrability can be investigated through a
modified Lax pair approach as introduced in [\Bowcock] to study 
boundary Toda theories.  We plan to address these questions in the future.


\zmsecspc
\noindent
{\zmsec Acknowledgments}\par
\nobreak
\zmtsecspc
We would like to thank P. Fendley, R. Konik, H. Saleur, P. Wiegmann
 and especially 
F. Lesage for their comments and suggestions.  This work is supported in
part by the National Science Foundation, in part through the 
National Young Investigator Program. \  Z.\ S.\ Bassi also acknowledges support
from the Olin Foundation.

\newpage

\equationnumber=0
\def\zenA{\the\equationnumber}
\def\zpA{
   \global\advance\equationnumber by 1 
   ({\rm A}\zenA)}%

\noindent
{\zmsec Appendix}\par
\nobreak
\zmtsecspc
We briefly discuss how the anisotropic spin 1/2 Kondo model can be mapped 
onto the one dimensional bosonic field theory
$$     H^K = \fb \zix \left( (\d t\phi)^2 + (\d x \phi)^2 \right) +
\l \left( S_+ e^{i\b\phi(0)/2} + S_- e^{-i\b\phi(0)/2} \right), \eqno\zpA $$
\edef\KfullhamZAa{A\zenA}%
where $\l$ and $\b$ are free parameters explained below.  A more complete
derivation can be found in [\Legett] (see also [\Guinea]), which 
we will follow closely.  
The map involves the technique of abelian bosonization, 
as was first applied to the Kondo model by Schotte [\Schotte].  
For further information on bosonization see [\Delft]-[\Emery] and 
references therein.

We begin with the anisotropic spin 1/2 Kondo Hamiltonian
$$ H^K = H^F_0 + H_{\rm int} \eqno\zpA $$
$$ H^F_0 = \sum_{\vk\s} \ep(\vk) c_{\vk\s}^\dg c_{\vk\s} \eqno\zpA $$
$$ H_{\rm int} =
{J_z\over 2} s_z \sum_{\vk\vk^\prime\s{\s^\prime}} c_{\vk\s}^\dg
{(\s^z)}_{\s{\s^\prime}}  c_{{\vk^\prime}{\s^\prime}} +
{\jp\over 2} \sum_{\vk\vk^\prime\s{\s^\prime}}\left(  s_x c_{\vk\s}^\dg
{(\s^x)}_{\s{\s^\prime}}  c_{{\vk^\prime}{\s^\prime}} + s_y c_{\vk\s}^\dg
{(\s^y)}_{\s{\s^\prime}}  c_{{\vk^\prime}{\s^\prime}} \right). 
\eqno\zpA $$
The interaction is taken to be pointlike with only s-wave scattering.
Thus the plane wave states can be expanded in spherical waves 
about the impurity, with only the angular momentum $l=m=0$ modes being 
retained.  Also since we are usually interested in low energy 
excitations near the fermi surface, $\ep(\vk)$ can be expanded as
$$ \ep(k) = \ep_F + v_F(k- k_F), \eqno\zpA $$
where $k =|\vk|$.  With these approximations, and measuring the
energy and momentum relative to the fermi surface, the free part
becomes (with $ v_F =1$)
$$ H^F_0 = \sum_{ p \s} p c_{p\s}^\dg c_{ p\s}, \eqno\zpA $$
where $c^\dg_{p\s}$ create s-wave electrons of momentum $k = k_F+p$. 
As is common for bosonization, in fact necessary [\Delft], we allow $p$ to be 
unbounded in $H_0^F$ and regularize by introducing an exponential cutoff 
in the interaction. The ground state consists of all 
states with $p< 0$ filled.  Using the fermion fields
$$ \ppt_\s^\dg(x) = {1\over\sqrt{L}} \sum_p c_{p\s}^\dg e^{-ip x},\eqno\zpA$$
where $L$ is the normalization length and 
$p= 2\pi n/L, n=0,\pm 1,\pm 2,\ldots$, combined with a redefinition of the
coupling constants, the interaction takes the form
$$ H_{\rm int} = {J_z\over 4} S_z \left(\ppt^\dg_\zu \ppt_\zu(0) -
\ppt^\dg_\zd \ppt_\zd(0)\right) + 
{\jp\over 2}\left(S_+\ppt^\dg_\zd \ppt_\zu(0)
+ S_- \ppt^\dg_\zu \ppt_\zd (0) \right),\eqno\zpA $$
where $S_z = \s^z$ and $S_\pm = {1\over 2}(\s^x \pm i\s^y)$.  The coupling 
constants are now dimensionless.

To bosonize, it is convenient to separate the so-called fermion charge
and spin degrees of freedom.  Define the following charge and spin density
operators for $k>0$
$$ \r_c(k) = \sum_p ( c_{p+k \zu}^\dg c_{p\zu} + c_{p+k\zd}^\dg c_{p\zd}),
\quad \r_c(-k) = \r_c^\dg(k) \eqno\zpA $$
$$ \r_s(k) = \sum_p ( c_{p+k \zu}^\dg c_{p\zu} - c_{p+k\zd}^\dg c_{p\zd}),
\quad \r_s(-k) = \r_s^\dg(k). \eqno\zpA $$
Then the operators
$$ a_k^c = \sqrt{{\pi\over k L}} \r_c(-k), \quad
a_k^s = \sqrt{{\pi\over k L}} \r_s(-k), \eqno\zpA $$
obey canonical boson commutation relations
$$ [a^c_k, a^{c\dg}_{k^\prime}] = [a^s_k, a^{s\dg}_{k^\prime}] =
\delta_{k k^\prime}, \eqno\zpA $$
when acting on the ground state.  Furthermore, instead of $H_0^F$ we can use
the boson Hamiltonian
$$ H_0^B = \sum_{k>0} k (a^{c\dg}_k a^c_k + a^{s\dg}_k a^s_k), \eqno\zpA$$
to obtain the same dynamics. The charge and spin sectors separate 
in the linear approximation.  The $J_z$ term can now be expressed as
$$  {J_z\over 4} S_z \left(\ppt^\dg_\zu \ppt_\zu(0) -
\ppt^\dg_\zd \ppt_\zd(0)\right) = {J_z\over 4} S_z
\sum_{k>0} e^{-ak/2} \sqrt{k\over \pi L} (a_k^s + a_k^{s\dg}). \eqno\zpA$$
Here we have introduced a cutoff which eliminates values of 
$k> a^{-1}\sim k_F$, thus regularizing divergent momentum sums.  To
write the $\jp$ interaction in terms of the boson degrees of
freedom, we make use of the ``bosonization" rule
$$ \ppt_{\zu,\zd}(x) = {1\over\sqrt{2\pi a}} e^{-i\pt_{\zu,\zd}(x)},
\eqno\zpA $$
\edef\KbdZA{A\zenA}%
where the fields $\pt_{\zu,\zd}$ are
$$ \pt_{\zu,\zd} = i \sum_{k>0} e^{-ak/2} \sqrt{2\pi\over k L}
\left( b_{k\zu,\zd} e^{-i k x} - b^\dg_{k\zu,\zd} e^{ikx} \right),
\quad \pt_{\zu,\zd}^\dg = \pt_{\zu,\zd} \eqno\zpA $$
with the $b$ operators related to the charge and spin operators
$$ b_{k\zu} = \fa(a_k^c + a_k^s),\quad b_{k\zd} = \fa(a_k^c - a_k^s).
\eqno\zpA$$
For the same spin index $\s$ $(\zu,\zd)$, the exponentials in (\KbdZA) 
behave as fermion fields.
The only problem with $\exp(\pm i\pt_\s)$ is that for different spins
$\s\neq\s^\prime$ these commute.  This can be fixed by using more complex
representations, such as including Klein factors [\Delft] (see 
also [\Emery],[\Mandel]).
The Klein factors serve to produce the correct commutation relations as 
well as create and destroy electrons, just as the fields $\ppt^\dg$ and
$\ppt$.  The exponentials alone cannot change the number of electrons.
We will not consider these important issues here.   With (\KbdZA) 
the $\jp$ term can be written as
$$ {\jp\over 2}\left(S_+\ppt^\dg_\zd \ppt_\zu(0)
+ S_- \ppt^\dg_\zu \ppt_\zd (0) \right) =
{\jp\over 4\pi a}\left( 
S_+ e^{i\sqrt{2\pi} \pt(0)} + S_- e^{-i\sqrt{2\pi} \pt(0)} \right),
\eqno\zpA$$
where 
$$ \pt(x) = {1\over \sqrt{2\pi} } \bl \pt_\zd(x) - \pt_\zu(x)\br
= -i\sum_{k>0} e^{-ak/2}\sqrt{{2\over kL}} \left( a_k^s e^{-ikx} - 
 a_k^{s\dg} e^{ikx} \right) \eqno\zpA $$
We see that the exchange interaction involves only the spin sector, the
charge sector is completely decoupled.  We will only be 
interested in the spin sector and drop the $s$ superscript.  Combining
everything we have
$$ H = \sum_{k>0} k a_k^\dg a_k - 
{J_z\over 4} {1\over \sqrt{2\pi}} S_z \d x \pt(0) + {\jp\over 4\pi a}\left(
S_+ e^{i\sqrt{2\pi} \pt(0)} + S_- e^{-i\sqrt{2\pi} \pt(0)} \right).
\eqno\zpA$$
\edef\KhamredZA{A\zenA}%

Now we can write down the boson field theory Hamiltonian associated
with (\KhamredZA).  Consider the following action on the half line
$$ S = \fb \zix \bl (\d t \phi)^2 - (\d x \phi)^2 \br. \eqno\zpA $$
\edef\KfreeboseactZA{A\zenA}%
Varying the action and requiring the boundary terms to vanish 
gives the boundary condition
$$ \d x \phi = 0,\ \ {\rm at \ }x=0.\eqno\zpA $$
\edef\KbdyeomZA{A\zenA}%
A mode expansion satisfying the bulk equation of motion,  
$ ({\d t}^2 - {\d x}^2)\phi = 0$, can be written as
$$ \phi(t,x) = -i \int_{-\infty}^\infty {dk\over\sqrt{2\pi}}\,
{1\over \sqrt{2 |k|}} \bl a(k) e^{-i\vk {\vec x}} - 
a^\dg(k) e^{i\vk{\vec x}}\br,\eqno\zpA$$
\edef\KbosemeZA{A\zenA}%
where $\vk \vx = |k|t-kx$.  The boundary condition (\KbdyeomZA)
requires that $a(k) = a(-k)$.  This allows us to separate $\phi$
into its left and right moving parts as follows
$$ \phi = \phi_L + \phi_R,\quad
\phi_{L,R}(t,x) = -i\int_0^{\infty} {dk\over\sqrt{2\pi}}\,
{1\over\sqrt{2 k}}\bl
a(k) e^{-ik(t\pm x)} - a^\dg(k) e^{ik(t\pm x)} \br.\eqno\zpA $$
Since $\phi_L(x) = \phi_R(-x)$, the free Hamiltonian obtained from 
(\KfreeboseactZA) is
$$ H^K_0 = \int_0^{\infty} dk\, k a^\dg(k) a(k),\eqno\zpA $$
which is just the continuum limit of the free part of (\KhamredZA).  
Also taking the continuum limit of $\pt$ we find that
$$ \pt(t,x=0) = \phi(t,0) = 2\phi_L(t,0) = 2\phi_R(t,0),\quad
\d x\pt(t,x=0) = \d t\phi(t,0)=\pi(t,0).\eqno\zpA $$
Of course we should include the exponential cutoff in (\KbosemeZA).  This
shows that the anisotropic Kondo model (to be exact, the spin sector) is
equivalent to the scalar boundary field theory  
$$ H^K = \fb\zix \bl (\d t \phi)^2 + (\d x\phi)^2 \br - {J_z\over 4} 
{1\over \sqrt{2\pi}} S_z \pi(0)  + {\jp\over 4\pi a}\bl
S_+ e^{i\sqrt{2\pi} \phi(0)} + S_- e^{-i\sqrt{2\pi} \phi(0)} \br.
\eqno\zpA$$

Finally to obtain (\KfullhamZAa) we make the unitary transformation
$$ H\rightarrow U H U^\dg, \quad U = e^{i\mu S_z \phi(0)},\eqno\zpA$$
where $\mu$ is a constant.  
The transformed Hamiltonian, apart from constants, is found to be
$$ H^K = \fb\zix \bl (\d t \phi)^2 + (\d x\phi)^2 \br - \bl{J_z\over 4} 
{1\over \sqrt{2\pi}}+ \mu\br S_z \pi(0)  + {\jp\over 4\pi a}\bl
S_+ e^{i(\sqrt{2\pi}+2\mu) \phi(0)} +S_- e^{-i(\sqrt{2\pi}+2\mu)\phi(0)} \br.
\eqno\zpA$$
Choosing $\mu$ to be
$$ \mu = - {J_z\over 4} {1\over \sqrt{2\pi}},\eqno\zpA $$
the $J_z$ term cancels, leaving the boundary field theory
 Hamiltonian  (\KfullhamZAa)
$$ H^K = \fb\zix \bl (\d t \phi)^2 + (\d x \phi)^2\br + \l \bl
S_+ e^{i\b\phi(0)/2} + S_- e^{-i\b\phi(0)/2} \br,
\eqno\zpA$$
\edef\KhamZA{A\zenA}%
where
$$ \l = {\jp \over 4\pi a}, \quad 
\b = \sqrt{2\pi} \bl 2 - {J_z\over 2\pi} \br. \eqno\zpA $$
The coupling $\l$ has a canonical dimension of mass 
whereas $\b$ is dimensionless.
For an antiferromagnetic $J_z$, $4\pi>J_z>0$, we have $0<\b<\sqrt{8\pi}$.
If we take $\jp$ to be small, then the isotropic Kondo model corresponds
to $J_z$ being small $J_z\approx \jp$.  Thus the isotropic point occurs
approximately at $\b=\sqrt{8 \pi}$.  We have shown that the anisotropic
Kondo model can be described by the bosonic boundary field theory (\KhamZA).

\newpage

\noindent
{\zmsec References}
\vskip 0.3 in

\def\adp{Adv.\ Phys.\ }
\def\anp{Ann.\ Phys.\ } 
\def\cmp{Commun.\ Math.\ Phys.\ }
\def\ijmpa{Int.\ J.\ Mod.\ Phys.\ A\ }

\def\jpc{J.\ Phys.\ C\ }
\def\jsp{J.\ Stat.\ Phys.\ }
\def\np{Nucl.\ Phys.\ }

\def\prb{Phys.\ Rev.\ B\ }
\def\prd{Phys.\ Rev.\ D\ }

\def\pla{Phys.\ Lett.\ A\ }
\def\plb{Phys.\ Lett.\ B\ }
\def\prl{Phys.\ Rev.\ Lett.\ }
\def\ptp{Prog.\ Theor.\ Phys.\ }
\def\rmp{Rev.\ Mod.\ Phys.\ }
\def\spj{Sov.\ Phys.\ JETP [Zh.\ Eksp.\ Teor.\ Fiz.]\ }
\def\spjl{Sov.\ Phys.\ JETP Lett.\ [Pis'ma Zh.\ Eksp.\ Teor.\ Fiz.]\ }
\def\zp{Z.\ Phys.\ }

\newdimen\zbibindent
\zbibindent=25pt
\def\box#1{\par\vskip5pt\noindent
\hangindent=\zbibindent
\hbox to \zbibindent{{\tt [#1]}\hfil}%
\ignorespaces}

\box{1}J. Kondo, \ptp {\bf 32} (1964) 37.

\box{2}K. G. Wilson, \rmp {\bf 47} (1975) 773.

\box{3}N. Andrei, \prl {\bf 45} (1980) 379.

\box{4}V. M. Filyov and P. B. Wiegmann, \pla {\bf 76} (1980) 283.

\box{5}P. B. Wiegmann, \spjl {\bf 31} (1980) 392.

\box{6}P. B. Wiegmann, \jpc {\bf 14} (1981) 1463.

\box{7}V. A. Fateev and P. B. Wiegmann, \pla {\bf 81} (1981) 179.

\box{8}N. Andrei, K. Furuya and J. H. Lowenstein, \rmp {\bf 55} (1983) 331.

\box{9}A. M. Tsvelick and P. B. Wiegmann, \adp {\bf 32} (1983) 453.

\box{10}I. Affleck, \np B {\bf 336} (1990) 517.

\box{11}I. Affleck and A. W. W. Ludwig, \np B {\bf 360} (1991) 641.

\box{12}I. Affleck, Acta Phys.\ Polon.\ B {\bf 26} (1995) 1869.

\box{13}P. Fendley, F. Lesage and H. Saleur, \jsp {\bf 85} (1996) 211. 

\box{14}P. Fendley and H. Saleur, \prl {\bf 75} (1995) 4492.

\box{15}Z. S. Bassi and A. LeClair, \prb {\bf 60} (1999) 615. 


\box{16}P. Fendley, H. Saleur and N. P. Warner, \np B {\bf 430} (1994) 577.

\box{17}F. Lesage, H. Saleur and S. Skorik, \prl {\bf 76} (1996) 3388.

\box{18}S. Ghoshal and A. B. Zamolodchikov, \ijmpa {\bf 9} (1994) 3841.

\box{19}E. Corrigan {\it et al.}, \plb {\bf 333} (1994) 83.

\box{20}A. LeClair, Ann.\ Phys.\ {\bf 271} (1999) 268. 


\box{21}M. Ameduri, R. Konik and A. LeClair, \plb {\bf 354} (1995) 376.

\box{22}P. Ginsparg, ``Applied Conformal Field Theory", in Les Houches
1988 Lectures, Eds.\ E. Br\'ezin and J.~Zinn-Justin, Elsevier Science
Publishers, 1989.

\box{23}S. Coleman, \prd {\bf 11} (1975) 2088.

\box{24}S. Mandelstam, \prd {\bf 11} (1975) 3026.

\box{25}D. Bernard and A. LeClair, \cmp {\bf 142} (1991) 99.

\box{26}P. B. Wiegmann and A. M. Finkelshtein, \spj {\bf 75} (1978) 204.

\box{27}P. Fendley, \prl {\bf 71} (1993) 2485.

\box{28}A. Fring and R. K{\" o}berle, \np B {\bf 421} (1994) 159.

\box{29}I. S. Gradshteyn and I. M. Ryzhik, {\it Table of Integrals, Series
and Products}, Academic Press, Inc., 1994.

\box{30}A. B. Zamolodchikov and Al. B. Zamolodchikov, \anp {\bf 120}
(1979) 253.

\box{31}A. LeClair and A. W. W. Ludwig, {\it Minimal Models with Integrable
Local Defects}, \hfil\break preprint hep-th/9708135 (ITP-97-080), to appear
in \np B.

\box{32}P. Bowcock {\it et al.}, \np B {\bf 445} (1995) 469.

\box{33}A. J. Leggett {\it et al.}, \rmp {\bf 59} (1987) 1. 

\box{34}F. Guinea, V. Hakim and A. Muramatsu, \prb {\bf 32} (1985) 4410.

\box{35}K. D. Schotte, \zp {\bf 230} (1970) 99.

\box{36}J. von Delft and H. Schoeller, Annalen der Physik {\bf 7} (1998) 225. 

\box{37}R. Shanker, ``Bosonization: How to make it work for you in Condensed 
Matter", Lectures given at the BCSPIN School, Katmandu, May 1991,
in {\it Condensed Matter and Particle Physics}, Eds.\ 
Y. Lu, J. Pati and Q. Shafi, World Scientific, 1993.

\box{38}M. Stone, {\it Bosonization}, World Scientific, 1994.

\box{39}V. J. Emery, ``Theory of the One-Dimensional Electron Gas" in
{\it Highly Conducting One-Dimensional Solids"}, Eds.\ J. T. Devreese,
R. P. Evrard and V. E. van Doren, Plenum, 1979.

\end